\documentclass[sigconf]{acmart}
\usepackage{booktabs} 
\usepackage{pbox}
\usepackage{csvsimple}
\usepackage{amssymb}
\usepackage{amsmath}
\usepackage{rotating}
\usepackage{multicol}
\usepackage{multirow}
\usepackage{xcolor}
\usepackage[export]{adjustbox}
\usepackage{enumitem}
\usepackage{array}
\usepackage{tabularx}
\usepackage{bigstrut}
\usepackage{subcaption}
\usepackage{graphicx}
\usepackage{pgfplots}
\usepackage{soul}
\usepackage{pgfplots}
\usepackage[breakable, theorems, skins]{tcolorbox}
\usepackage{marvosym}
\usepackage[flushleft]{threeparttable}
\pgfplotsset{compat=1.3}
\usepackage{color, colortbl}
\usepackage{tikz}
\usepackage{pgf}
\usepackage{balance}
\usetikzlibrary{positioning}
\usepackage[linesnumbered,ruled,vlined]{algorithm2e}
\usetikzlibrary{arrows}
\usepackage[flushleft]{threeparttable}
\graphicspath{ {images/} }

\sethlcolor{lightgray}

\newcolumntype{R}[2]{%
>{\adjustbox{angle=#1,lap=\width-(#2)}\bgroup}%
l%
<{\egroup}%
}

\usetikzlibrary{patterns}
\usetikzlibrary{arrows}
\pgfdeclarepatternformonly{north east lines 
wide}{\pgfqpoint{-1pt}{-1pt}}{\pgfqpoint{10pt}{10pt}}{\pgfqpoint{9pt}{9pt}}%
{
\pgfsetlinewidth{0.4pt}
\pgfpathmoveto{\pgfqpoint{0pt}{0pt}}
\pgfpathlineto{\pgfqpoint{9.1pt}{9.1pt}}
\pgfusepath{stroke}
}

\definecolor{Gray}{gray}{0.9}

\newcommand{\bi}{\begin{itemize}}
\newcommand{\ei}{\end{itemize}}
\newcommand{\be}{\begin{enumerate}}
\newcommand{\ee}{\end{enumerate}}
\newcommand{\tion}[1]{\S\ref{sect:#1}}
\newcommand{\fig}[1]{Figure~\ref{fig:#1}}

\newcommand{\eq}[1]{Equation~\ref{eq:#1}}

\tikzset{%
body/.style={inner sep=0pt,outer 
sep=0pt,shape=rectangle,draw,thick,pattern=north east lines wide},
dimen/.style={<->,>=latex,thin,every rectangle node/.style={fill=white, 
midway,font=\small}},
symmetry/.style={dashed,thin},
}

\usepackage[leftmargin=2em,rightmargin=2em]{quoting}

\acmConference[ICSE'18]{International Conference on Software Engineering}{May 
2018}{Gothenburg, Sweden}

\begin{document}

\title{What is the Connection Between Issues, Bugs, and Enhancements?
(Lessons Learned
from
800+ Software Projects)}
\author{
Rahul Krishna, Amritanshu Agrawal, Akond Rahman, 
Alexander Sobran*, and Tim Menzies}
\affiliation{North Carolina State University, *IBM Corp} 
\email{i.m.ralk@gmail.com, [aagrawa8, aarahman]@ncsu.edu, asobran@us.ibm.com, 
tim@menzies.us}


\begin{abstract}
Agile teams juggle multiple tasks so professionals are often assigned to multiple projects, especially in service organizations that monitor and maintain a large suite of software for a large user base. 
If we could predict changes in
project conditions
changes, then managers could  better adjust the
staff allocated to those projects.

This paper builds such a predictor using data from  832  open source and 
proprietary applications. 
Using a time series analysis of the   last 4 months
of {\em issues}, we can forecast how many {\em bug reports} and \textit{enhancement requests} will be generated  next month. 

The forecasts made in this way only require a frequency count of this issue reports 
(and do \underline{not} require an historical record of bugs found in the project). That is, this kind of predictive model is very easy to deploy within a project. We hence strongly recommend this method for forecasting future issues, enhancements, and bugs in a project. 
\end{abstract}



\maketitle


\pagestyle{plain}
\section{Introduction}
\label{sect:intro}

In the early days of software engineering, when doing any single project
was a Herculean task,   developers were often assigned to one project for months at  a time.
In the age of agile~\cite{abrahamsson2017agile, 
dybaa2008empirical, misra2009identifying, begel2007usage}, that has changed. How should management practices change to better accommodate agile developments?

One principle of 
agile  is to use multitasking to get work done.
Agile teams juggle multiple tasks so professionals are often    assigned to multiple projects, especially in service organizations that monitor and maintain a large suite of software for a large user base.

When project conditions
changes, it would be very useful 
if managers can  adjust the
staff allocated to those projects.
Ideally, managers   overseeing multiple projects
would like to tell whether more/less developers will be required in the upcoming month in order to make informed choices about changes to staffing allocations.

This  ideal scenario might be achievable, given access
to enough   software projects.
For example, here we apply a  time series analysis to   832 proprietary and 
open source projects from GitHub repositories.  Data was gathered by sampling projects every week for an 
average period of two years. The mined data contained the following attributes: issue reports, bugs reports, and enhancement requests for each week. Trends in this data were
modeled using
\textit{\underline{A}uto\underline{R}egressive \underline{I}ntegrated
\underline{M}oving 
\underline{A}verage} (ARIMA)~\cite{box2015time}
(this is a standard method that has shown to  outperform other models 
such as linear regression or random walk~\cite{kenmei2008trend, ho1998use, pai2006software, amin2013approach, 
burtschy1997improving}). 
Using this data, we offer the following contributions:
\be
\item In hundreds of software projects, we prove the existence of  simple and predictable trends in issues, bugs and enhancements.
\item These different trends are closely associated. Hence, using past issue reports we can forecast (a)~future bugs and (b)~future enhancement requests.
\item
The forecasts made in this way only require  a frequency count of this issue reports. They do not require an historical record of bugs found in the project. That is, this kind of predictive model is very easy to deploy within a project.
\item
In studies with 832 times series models generated separately for all our projects, we show that the
 forecasts made in this way are remarkably accurate.
\ee
The rest of this paper is structured as follows. 
 After some notes on the    motivation of this work, this paper's   research questions are presented  in \S\ref{mr}.   \tion{related}  discusses some   related work.   \tion{method} presents our experimental methodology. In 
\tion{results}, we answer our research questions and discuss  lessons learned. In \tion{discuss} we discuss the practical implications of our results. In \tion{threats}, we present our threats to validity.  
 \tion{conclusion} presents   conclusions and directions for future work.

Before beginning, we digress to comment that this paper focuses on ``near-term forecasts''.
That is,    given the last 4 months of data from one project, we infer what is expected in the next month for that project. A bigger question would be ``for multiple projects, how to predict for many months into the future?''. While we have some  preliminary results on that bigger question, there are no definitive results to report at this time.  Accordingly, this paper reports
only near-term forecast results.

\subsection{Motivation and Research Questions}
\label{mr}
Why build yet another bug predictor? Why learn predictors for future bug reports using
data from prior issues? In the literature there are  any number bug prediction methods 
methods as witnessed by the plethora of papers in that area~\cite{krishna2017less, krishna2016too, menzies2007data, turhan2009relative, menzies2010defect}. 

For large cloud-based environments of service 
organizations supporting multiple languages and tools, we found that those 
methods had certain significant drawbacks.
Firstly, before anyone can use past bugs to forecast for future bugs,
{\em they need access to past bugs}.  For this, they could use APIs provided by platforms like GitHub to obtain temporal logs issues. However, this is not sufficient. After mining issues, these need to be carefully curated to identify the bugs. We spent over two months at IBM to manually categorize logs of issues into bugs and enhancements. Due to the significant amount of time and effort required to do this, in this paper we explore what can be achieved {\em with just} logs of issues.

Secondly, there was the problem of commissioning standard defect predictors for 
dozens of programming languages\footnote{Popular languages include Java, 
Python, 
Javascript, C++, Lua, Perl, Ruby, etc.}. Nearly all 
the prior defect prediction work in software engineering have placed focus only 
on a few languages like C++ and Java. Certainly, we could 
build our own but predictors, but merely building them is not the issue. Far more problematic 
is the issue of certifying that they work against known baseline data (which 
may be missing). In addition to that, maintaining all that software over all 
those languages would turn into an arduous task in itself.

Problems like the above forced us to consider radical alternatives to 
traditional defect prediction technology. In meetings to discuss those 
alternatives, we came across a simple alternative --- forecast bugs by looking 
at trends of past issue reports.
 We found that 
by \textit{mining only for issues}, we can construct accurate models that 
forecast for bugs and enhancements. To show that, this paper explores the following research questions.

\subsection*{RQ1: Are there temporal trends in our data?}

\begin{itemize}[leftmargin=-1pt]
	
	\item[] \textit{\textbf{Motivation:}} The first research question seeks to establish the existence of temporal trends in the attribute we have mined (i.e., issues, bugs, and enhancements). To assert this, we ask if the past temporal data of attributes mined can be used to construct time series models that forecast future trends for the same attributes.
	
	\item[] \textit{\textbf{Approach:}} For each of our 832 proprietary and 
	opensource 
	projects, we use the mined attributes (issues, bugs, and enhancements) and for each of these attributes we 
	construct a time series model with ARIMA. Then, we used these ARIMA models 
	to forecast future issues, bugs, and enhancements (Note: this is different from RQ3, there we 
	built an ARIMA model only on issues and used that to forecast for bugs and 
	enhancements).\\[-.2cm]
	
\end{itemize}

\noindent\begin{minipage}{\linewidth}
	\begin{center}
		\begin{tabular}{p{0.95\linewidth}}
			\arrayrulecolor{Gray}
			\hline
			 
			\rowcolor{Gray} \textit{\textbf{Result:}} ARIMA models built on past 
	temporal data of issues, bugs, and enhancements can be very accurate for 
	forecasting future values.\\\hline
		\end{tabular}
	\end{center}
\end{minipage}\bigstrut[t]\\[-.2cm]

\subsection*{\bf{RQ2: Are there correlations between mined attributes?}}

\begin{itemize}[leftmargin=-1pt]

\item[] \textit{\textbf{Motivation:}} Our second research question follows the 
report 
by Ayari et al.~\cite{ayari2007threats} regarding the correlation of issues 
reports with bugs and enhancement. Here we seek to establish this 
on our dataset of 832 projects. We ask if the time series trends of 
issues, bugs, and enhancements are correlated to each other. A strong 
correlation between these attributes would enable use to make use of models 
built on one attribute such as issues to forecast for other attributes.

\item[] \textit{\textbf{Approach:}} In each of our 832 proprietary and 
opensource 
projects, we compute the Spearman's $\rho$ value between pairs of 
attributes. 
A value close to 1 would indicate a strong positive correlation, a value 
close to -1 would indicate a strong negative correlation, and a value close 
to 0 would indicate no correlation.\\[-.3cm]
\end{itemize}

\vspace{0.1cm}
\noindent\begin{minipage}{\linewidth}
	\begin{center}
		\begin{tabular}{p{0.95\linewidth}}
			\arrayrulecolor{Gray}
			\hline
			 
			\rowcolor{Gray}  \textit{\textbf{Result:}} In proprietary projects certain 
pairs 
of attributes such as $\left< issues, bugs\right>$ and $\left< 
issues, 
enhancements\right>$ have relative a strong correlation. In opensource 
projects, on the other hand, the correlations between project attributes 
still exist but they are relatively weaker in comparison to proprietary 
projects.\\\hline
		\end{tabular}
	\end{center}
\end{minipage}\bigstrut[t]\\[-0.35cm]

\subsection*{RQ3: Can issue reports forecast for future bugs and enhancements?}
\begin{itemize}[leftmargin=-1pt]

\item[] \textit{\textbf{Motivation:}} This research question naturally follows 
RQ2. 
Here we ask if it is possible to use time series models built on one attribute 
such as issues to estimate for another attributes such as bugs and enhancements.

\item[] \textit{\textbf{Approach:}} We construct an ARIMA model on time series 
data of 
issue reports for each project to forecast for bugs and enhancements. That 
is, we transfer ARIMA models between:
\begin{enumerate}
\item[] a) issues $\rightarrow$ bugs, and 	
\item[] b) issues $\rightarrow$ enhancements.
\end{enumerate} 
Then we compare the forecast values with the actual values by measuring the 
magnitude of residual error.\\[-0.4cm]

\end{itemize}

\vspace{1mm}
\noindent\begin{minipage}{\linewidth}
	\begin{center}
		\begin{tabular}{p{0.95\linewidth}}
			\arrayrulecolor{Gray}
			\hline
			 
			\rowcolor{Gray}   \textit{\textbf{Result:}} ARIMA models built on issues 
can be 
very accurate for forecasting future bugs and enhancements.\\\hline
		\end{tabular}
	\end{center}
\end{minipage}\bigstrut[t]\\[-0.4cm]

\subsection*{RQ4: Are the forecasts using issues better than with using past temporal data?}
\begin{itemize}[leftmargin=-1pt]

\item[] \textit{\textbf{Motivation:}} In the final research question, we 
compare the 
errors between using time series model built with issues to 
forecast for future values of bugs and enhancements with forecasts 
using past temporal data of bugs and enhancements. As mentioned previously, it took us a significant amount of time to curate issue reports into bugs and enhancements. If the errors in using only issues for forecast is statistically comparable to using each time series separately, then we can establish that the time series trend of issues can indeed forecast for bugs and enhancements and that may save a lot of time and effort.

\item[] \textit{\textbf{Approach:}} As before, for each of our 832 proprietary 
and 
opensource 
projects, we construct two time series models with ARIMA:
\be
\item[] a) $\mathit{ISSUE}$ : ARIMA model built using past issue report trend.
\item[] b) $\mathit{LOCAL}$ : ARIMA model built using past bug report trend and 
past enhancement request trend.
\ee

We use both of these models to forecast for future bugs and enhancements. 
Then, we compute the error in forecasts with   $\mathit{ISSUE}$ and 
$\mathit{LOCAL}$. Finally, we use a statistical test (Welch's t-test) to 
compare the errors.
 
\end{itemize}

\vspace{1mm}
\noindent\begin{minipage}{\linewidth}
	\begin{center}
		\begin{tabular}{p{0.95\linewidth}}
			\arrayrulecolor{Gray}
			\hline
			 
			\rowcolor{Gray}   \textit{\textbf{Result:}} Forecast errors of 
$\mathit{ISSUE}$ are statistically similar to $\mathit{LOCAL}$. That is, we can avoid all the complexity of bug mining by just building times series models from issue reports. \\\hline
		\end{tabular}
	\end{center}
\end{minipage}\bigstrut[t]

\section{Related Work}
\label{sect:related}

The study of evolution of software systems over time has been subject to much research over the past decades. Several researchers have attempted to model long term temporal behavior of aspects of software systems such as structural changes, line of code, etc.~\cite{godfrey2001growth, lehman1985program}.  Godfrey and Qiang studied the growth of a number of opensource systems such as linux and gcc compilers. For linux, they report that the growth rate is geometric. Using this, they were able to develop a time series approach to model long term growth of such systems. Such time series models have been very popular with several software engineering researchers~\cite{lehman1985program, fuentetaja2002software, turski1996reference, wu2006seeking}. Fuentetaja and Bagert~\cite{fuentetaja2002software} used time series growth models of software systems to forecast how much memory systems may use. They demonstrate that these time series growth models of software systems exhibit a power law. Using a Detrended Fluctuation Analysis method they were able to establish a theoretical basis for some trends they noticed in software evolution. Wu et al.~\cite{wu2006seeking} studied the existence of correlations in time series over long stretches of time. Using a Re-scaled Range Analysis technique~\cite{hurst1951long}, they reported the existence of temporal signature in the software systems and that these systems exhibit a macroscopic behaviour of self-organized criticality.

Another area of software engineering that has seen extensive use of time series 
modeling is software reliability. Several researchers, from as early as
1972~\cite{jel72}, have attempted to describe time series models for measuring 
software reliability. This has resulted in several 100 different forms 
of time series models~\cite{Lyu07}. These initial models made strong stochastic 
assumptions about the data and were grouped in to two categories: (1) Failure 
interval models~\cite{jel72, goel85}, and (2) Failure count 
models~\cite{goel79, lyu96}. However, these models made several 
unrealistic assumptions such as independence of time between failures, 
immediate correction of detected faults, and correcting faults without 
introducing new faults~\cite{goel85, Zeitler91, wood97}. Zeitler et al.~\cite{Zeitler91} cautioned that this was a major impediment since, the real world use of these models was not practical. 

In response to the above, researchers explored approaches based on 
non-parametric statistics~\cite{Robinson87, Barghout98} and Bayesian networks as possible solutions~\cite{Neil96, Bai05, Fenton08, 
Wiper12}. 
However, even though these non-parametric
approaches are able to address the unrealistic assumptions issue;
they cannot completely address the applicability and predictability
issues. As a result, other methods based on neural networks and other
machine learning methods~\cite{karunanithi92, Pai06, Kiran07, Zaidi08, Lo09, 
Yang10, Moura11} were introduced. However, the issue with these approaches is that they 
require a large training data set as input/output examples to obtain 
accurate forecasts, which was computationally intensive and time consuming 
process. 

An  alternative to the above approaches was offered by Zeitler et 
al.~\cite{Zeitler91}; they recommended the use of time series models such as 
ARIMA. They offer strong statistical evidence to show the time series models 
(especially ARIMA) are best suited for mapping system failures over 
time~\cite{Zeitler91}. As a result, a number of researchers have applied time 
series models, especially ARIMA models~\cite{Singpur85, Chatterjee97, Xie99, 
Junhong05, amin2013approach}. These researchers have shown that ARIMA models 
have  the ability to give accurate forecasts~\cite{amin2013approach}. Yuen et 
al.~\cite{chong1988analyzing} used ARIMA models to predict the evolution in the 
maintenance phase 
of a software project with sampling
periods of one month. Kemerer et al.~\cite{kemerer1999empirical} use ARIMA 
models
to predict the monthly number of changes of a software
project. Herraiz et al.~\cite{herraiz2007forecasting} used ARIMA models to 
model time series changes in Eclipse IDE by smoothing using kernel methods. 

Although ARIMA models have been well established for time series analysis, our reading of the literature has indicated that many of these methods do not perform a comprehensive empirical study to establish it's usefulness. The 
success of ARIMA models from Ayman et al.~\cite{amin2013approach} for instance 
was 
shown to work on 16 projects (some as small as 5000 LOC). Similarly, Kenmei et al.~\cite{kenmei2008trend} et al. performed their analytics on only 3 software systems, Eclipse, Mozilla, and JBoss.  Our paper overcomes this limitation 
by performing a large-scale case study with the ARIMA model on 832 open source 
and 
proprietary applications. In doing so, we demonstrate promising results showing that it is possible to mine projects on GitHub over long stretches of time to generate 
time series models which in turn can be used to forecast the number 
of issues, bugs, and enhancements.



\section{Methodology}
\label{sect:method}	

This section first details our datasets and our policy for gathering and filtering these datasets in \tion{datasets}. Then, we discuss time 
series modeling with ARIMA in \tion{ARIMA}. After that, we discuss the proposed 
forecasting approach and statistical measures in \tion{approach} and 
\tion{stats} respectively.

\subsection{Datasets}
\label{sect:datasets}

To perform our experiments we use open-source projects from GitHub, and 
proprietary projects obtained from our industrial partners at IBM Raleigh. These totaled to 1,646 different projects with 1,108 opensource projects and 538 proprietary projects. Our data selection strategy is as follows:

\be
\item In case of open source projects we select public GitHub projects that are included as a `GitHub showcase  project'. Of the publicly available projects hosted on GitHub, a selected set of projects are marked as `showcases', to demonstrate how a project can be 
developed in certain domain such as game development, music, etc.~\cite{gh:showcase}. Our assumption is that by selecting these GitHub projects we can start with a representative set of open source projects that enjoy popularity, and provide good examples of software development. Example of popular projects included in the GitHub showcase that we use for our analysis are: Javascript libraries such as `AngularJS'\footnote{https://GitHub.com/angular/angular.js} and  `npm'\footnote{https://GitHub.com/npm/npm}, and programming languages such as  `Go'\footnote{https://GitHub.com/golang/go}, `Rust'\footnote{https://GitHub.com/rust-lang/rust}, and `Scala'\footnote{https://GitHub.com/scala/scala}. For more examples, see \fig{sample_datasets}.

\item In case of proprietary projects our collaborating company (IBM) provided us a list of projects that are hosted on their private GitHub. We mine open source and proprietary projects, respectively, by using the public GitHub API, and a private API maintained by our collaborating company.

\ee

\begin{figure}[t!]
	\centering
	\includegraphics[width=\linewidth]{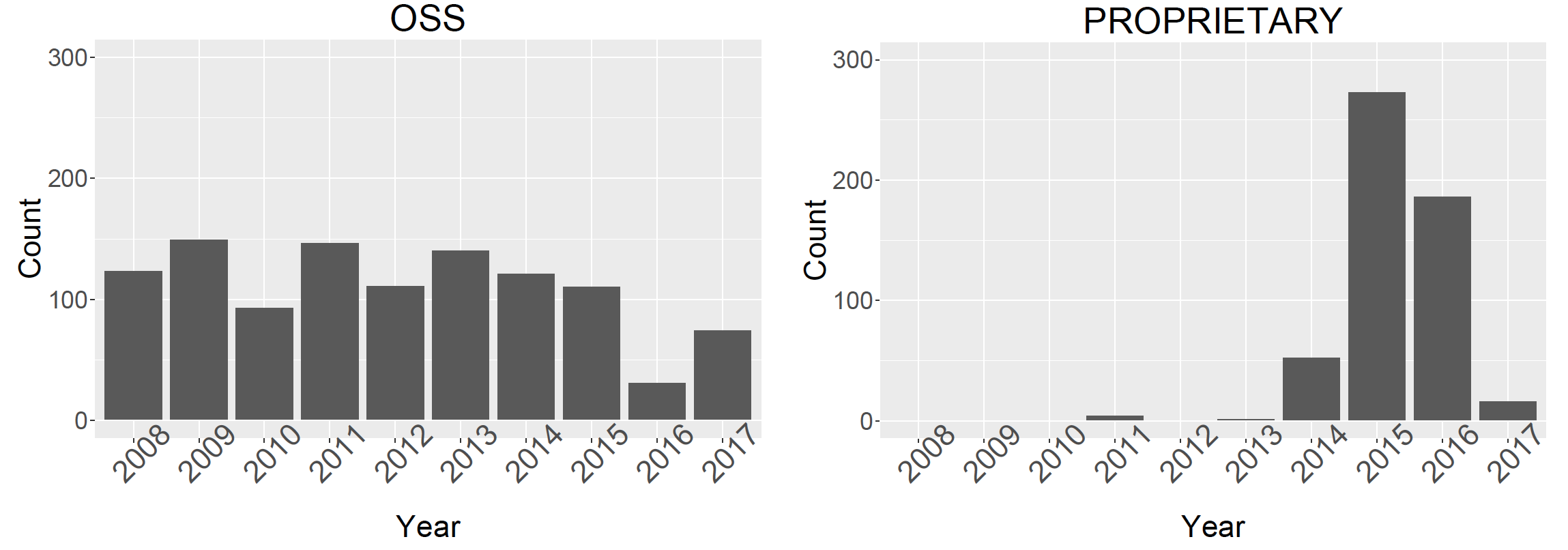}
	\caption{The project count and their start years. On the left, we plot the counts of open source projects started in each year. On the right, we count of proprietary projects started in each year.}
	\label{fig:start_date}
\end{figure}

Note that all the projects are hosted on GitHub. They have different start dates. We show the start dates of the proprietary and open source projects in \fig{start_date}. It is worth noting that a majority of these projects have a history of at least one year.

\subsubsection{Extracting Relevant Projects}

It is important to note that projects hosted on GitHub gives researchers a tremendous opportunity to extract necessary project information such as issues, bugs, and 
enhancements~\cite{kalliamvakou2014promises}~\cite{bird2009promises}~\cite{MunaiahCuration2017}. Unfortunately, it is possible that many of these projects can contain very short development activity, can be used for personal use, or not be related to software development at all~\cite{kalliamvakou2014promises}~\cite{bird2009promises}. These projects may bias our findings. Hence, we implement a set of rules to identify and discard these projects. We call these set of rules ``filters'' and they are designed such that only the projects that contain sufficient software development data for analysis pass this filter. 

As the first step of filtering, we identify projects that contain sufficient software 
development information using the following criteria. These criteria address 
the limitations of mining GitHub projects as highlighted by prior 
researchers~\cite{kalliamvakou2014promises}~\cite{bird2009promises}. The rest 
of the rules are listed below:

\begin{figure}[t!]
	\centering
	\includegraphics[width=0.8\linewidth]{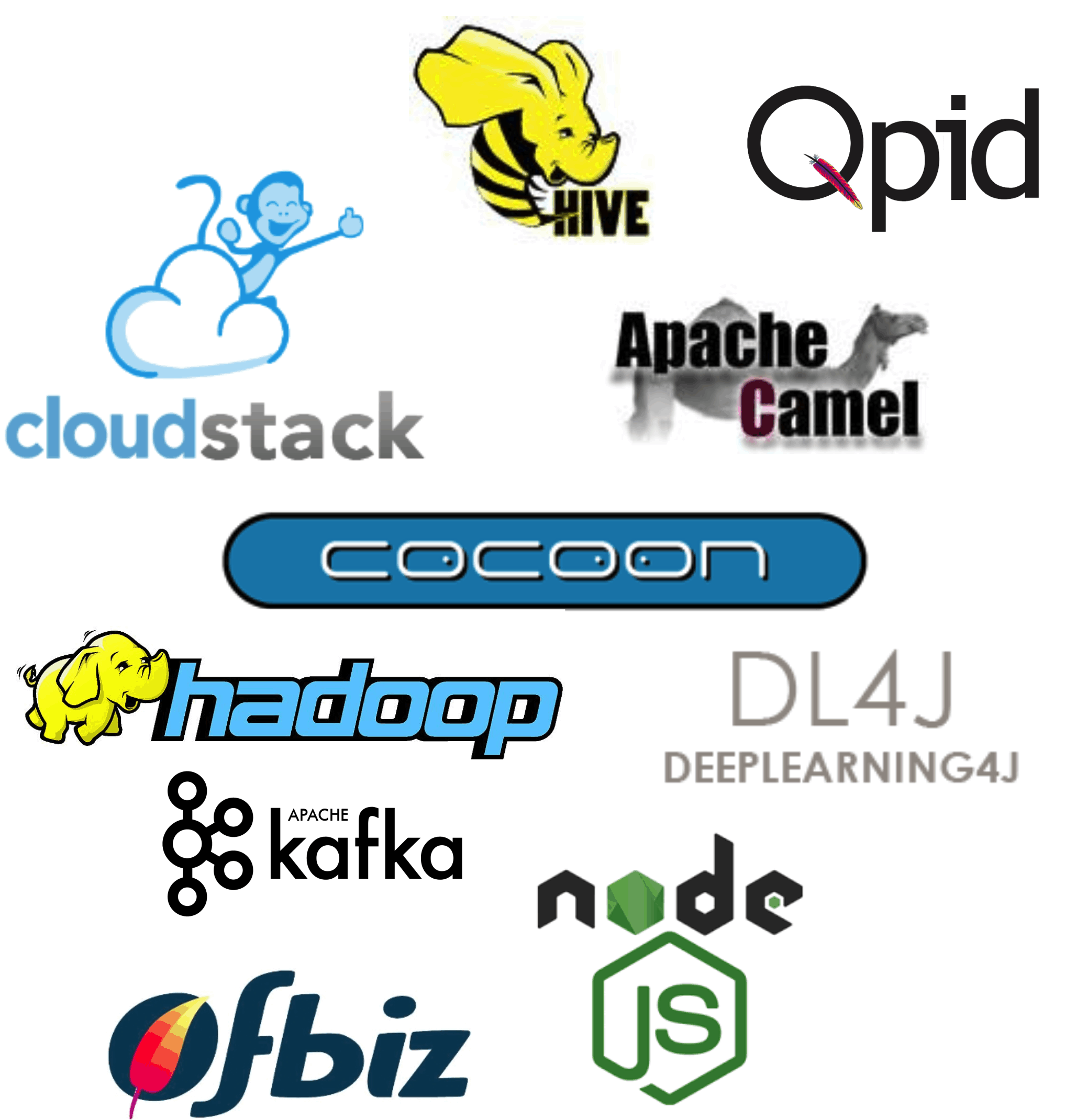}
	\caption{A subset of open source projects used in this study. In addition to these, our datasets contain a total of 1108 opensource projects and 538 proprietary projects.}
	\label{fig:sample_datasets}
\end{figure}

\begin{figure}[t!]
\footnotesize
\centering
\resizebox*{\linewidth}{!}{
\begin{tabular}{lrr}
\bottomrule
\multicolumn{1}{c}{\multirow{2}{*}{Sanity check}} & 
\multicolumn{2}{c}{Discarded project count}                     \\ 
\cmidrule{2-3} 
& \multicolumn{1}{l}{Proprietary} & \multicolumn{1}{l}{open-source} \\ 
\midrule
Collaboration (Pull requests $> 0$)      & 35                               
& 54                       \\
Commits $> 20$                           & 68                               
& 96                       \\
Duration $> 1$ year                      & 12                               
& 46                       \\
Issues  $> 10$                           & 60                               
& 89                       \\
Personal purpose (\# programmers $> 8$)  & 47                               
& 67                       \\
Releases \textgreater 0                  & 136                              
& 44                       \\
S\/W development only                    & 9                                
& 51                       \\ 
\midrule
Projects after filtering                 & 171                           & 
661                       \\ 
\bottomrule
\end{tabular}}
\caption{Count of projects that pass the filter. Upon completion, we are left 
with 171 proprietary and 661 open-source projects.}
\label{fig:sanity}
\end{figure}

\begin{itemize}[leftmargin=*]
\item{\textit{Collaboration}: Number of pulls requests are indicative of 
collaboration, and the project must have at least one pull request.}
\item{\textit{Commits}: The project must contain more than 20 commits.}
\item{\textit{Duration}:The project must contain software development 
activity of at least 50 weeks.}
\item{\textit{Issues}:  The project must contain more than 10 issues.}
\item{\textit{Personal Purpose}:The project must not be used and maintained 
by one person. The project must have at least eight contributors.}
\item{\textit{Releases}:The project must have at least one release.}
\item{\textit{Software Development}:The project must only be a placeholder 
for software development source code.}
\end{itemize}

After applying the aforementioned filter, from our initial pool of 1,108 open source projects and 538 proprietary we are left with 661 open source and 171 
proprietary projects. For details of how many projects were passed our each of our filter rules see~\fig{sanity}.  

From \fig{sanity} we observe that 59.6\% of the GitHub showcase projects pass the 
recommended sanity checks by researchers. The 447 projects filtered by 
applying the filter further emphasizes the need to validate software project 
data mined from GitHub before use lest they skew the findings.

\subsection{Time Series Modeling}
\label{sect:ARIMA}

Autoregressive Integrated Moving Average (ARIMA) models were proposed by Box 
and Jenkins~\cite{box2015time} in 1976. They
are now most commonly used to model time series data to forecast the future 
values. The ARIMA model extends ARMA (Autoregressive Moving Average) model by 
allowing for non-stationary time series to be modeled, i.e., a time series 
whose statistical properties such as mean, variance, etc. are not constant 
over time. 

A time series is said to be autoregressive moving average (ARMA) in nature 
with parameters $(p,q)$, if it takes the following form:
\begin{equation}
\label{eq:arima}
y_t=\sum_{i=1}^{p}\phi_i 
y_{t-i}+\sum_{i=1}^{q}\theta_i\epsilon_{t-i}+\epsilon_t
\end{equation}

\begin{figure}[t!]
	\centering
	\includegraphics[width=\linewidth]{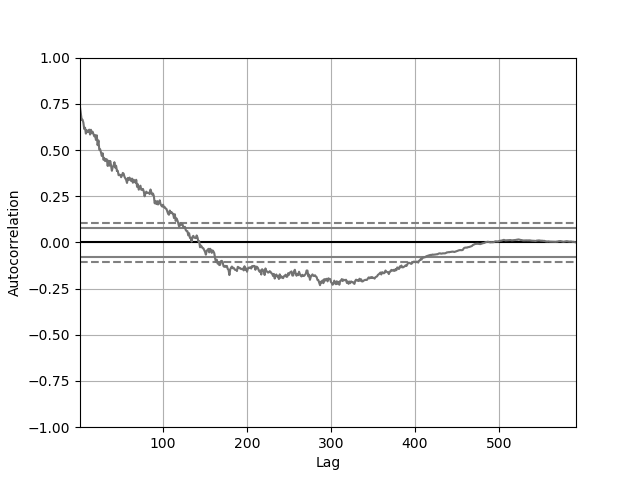}
	\caption{Autocorrelation plot of issue reports. The ``lag'' parameter 
	indicates the number of weeks over which the autocorrelation was computed. 
	Note that autocorrelation is significantly large for lag$\leq$20 weeks.}
	\label{fig:autocorr}
\end{figure}

Where $y_t$ is the current stationary observation, $y_{t-i}$ for \mbox{$i = 1, . . 
., p$}
are the past stationary observations, $\epsilon_t$ is the current error, and 
$\epsilon_{t-i}$
for $i = 1, . . ., q$ are the past errors. If this original time series 
$\{z_t\}$ is non-stationary, then $d$ differences can be done to transform it 
into a stationary one $\{y_t\}$. These differences can be viewed as a 
transformation denoted by $y_t = \triangledown^dz_t$, where 
$\triangledown^d=(1-B)^d$ where $B$ is known as a backshift operator. When 
this differencing operation is performed, it converts an ARMA (Autoregressive 
Moving Average) model into an ARIMA (Autoregressive Moving Integrated Average) 
model with parameters $(p,q,d)$.

Before using ARIMA, the observed time series has to be analyzed to 
select the parameters for $ARIMA(p,q,d)$. This requires the identification of 
the p, d, and q parameters. To do this, we take the following steps
\be
\item \textit{Estimating $p$:} The value of $p$ can be estimated by analyzing 
the \textit{autocorrelation} plot of the time series. This is a plot of 
correlation between the time series and the same time series separated by a 
given interval (called $lag$). To demonstrate this procedure, consider an 
example autocorrelation plot of the ArangoDB project in~\fig{autocorr}. In this 
figure, we see that the autocorrelation is significantly large for values of 
$lag\leq20$. So, we may set any $p<20$ for a good model. 
\item \textit{Estimating $d$:} The value of d has to be set taking into account 
whether the time series is stationary or not. If the time series is stationary 
the we may set $d = 0$, otherwise we set $d > 0$. To determine if a time series 
is stationary, we use the Dickey and Fuller test~\cite{dickey1979distribution}.
\item \textit{Estimating $q$:} The value of q can be set taking into account 
whether the time series measurements have error in measurements. In our case, 
since we mine the GitHub repositories with their official API, there are no 
measurement errors. Thus, we set $q = 0$.
\ee

After the ARIMA model has been constructed, we need to ensure that the given 
data can be accurately model by it. Like most time series modeling techniques, 
ARIMA has some inherent assumptions. It is therefore required that the data be 
preprocessed and the assumptions be satisfied before the ARIMA model is applied 
lest we risk inaccurate forecast. In the following, we list these preprocessing 
steps taken by us:
\be

\item \textit{Ensuring Normality:} ARIMA model assumes that the given time series data 
is approximately normally distributed. In order to ensure this, the data was transformed to approximate normal distribution using power transformations.

\item \textit{Ensuring Stationarity:} It is assumed that the time series has a constant 
mean, variance, and no trend over time. We use the Dickey and Fuller test~\cite{dickey1979distribution} to test for stationarity. If we note that the series is non-stationary, we transform using using differences. This can be achieved by setting appropriate $d$ value in $ARIMA(p,d,q)$.
\ee

Investigating whether the given time series data satisfies these assumptions 
is a critical task, because falling to satisfy the assumptions leads to 
selecting incorrect ARIMA model. In this work, we independently verified these 
assumptions and applied the necessary corrective transformations before applying the ARIMA model. Additionally, we performed extensive empirical evaluations to determine the values of $p,~q,\text{ and }d$ to be used in $ARIMA(p,q,d)$.

\subsection{Measuring Forecast Error}
\label{sect:stats}

To evaluate the quality of the ARIMA models used for forecasting, we compute the mean absolute error (MAE). MAE has been a preferred method to evaluate errors in time series analysis by researchers in several areas~\cite{willmott2005advantages,willmott2006use}. MAE is a measure of difference between two continuous variables. Assume X and Y are variables of paired observations that express the same phenomenon. Examples of Y versus X include comparisons of predicted versus observed, subsequent time versus initial time, and one technique of measurement versus an alternative technique of measurement. Consider a scatter plot of n points, where point i has coordinates (xi, yi). Mean Absolute Error (MAE) is the average vertical distance between each point and the Y=X line, which is also known as the One-to-One line. MAE is also the average horizontal distance between each point and the Y=X line.
The Mean Absolute Error is given by:
\begin{equation}
\label{eq:mae} 
MAE = \sum_{i=1}^{N}P_i\left|\hat{Y_i}-Y_{i}\right|=\sum _{i=1}^{N}P_i\left|e_{i}\right|
\end{equation}

Here, N represents the total number of unique values of issues, bugs, enhancements, etc. $P_i$ represents the probability of appearance of each of the unique values of issues, bugs, enhancements, etc. $Y_i$ denotes the actual value, $\hat{Y_i}$ denotes the predicted values. According to this formulation, lower values of MAE are considered to be better. 

Some researchers~\cite{amin2013approach} have endorsed the use other metrics 
such as mean squared error (MSE) or mean magnitude of relative error (MMRE) to 
measure the quality of forecast in time series models. We, however, refrain 
from using these measures for the following reasons:

\bi
\item MAE has a clear interpretation as the average absolute difference between $Y_i$ and $\hat{Y_i}$. Many researchers~\cite{willmott2005advantages,willmott2006use} find this measure desirable because its interpretation is clear. However, researchers frequently compute and misinterpret the Root Mean Squared Error (RMSE), which is not the average absolute error~\cite{willmott2005advantages,willmott2006use}.
\item In cases where the true value is close to or equal to zero, measures like MMRE fail to provide a accurate description of the error. When true values are zero, MMRE is extremely large; this skews the errors and risks leading to spurious interpretation of the errors. 
\ei

\begin{figure}
	\centering
	\includegraphics[width=0.7\linewidth]{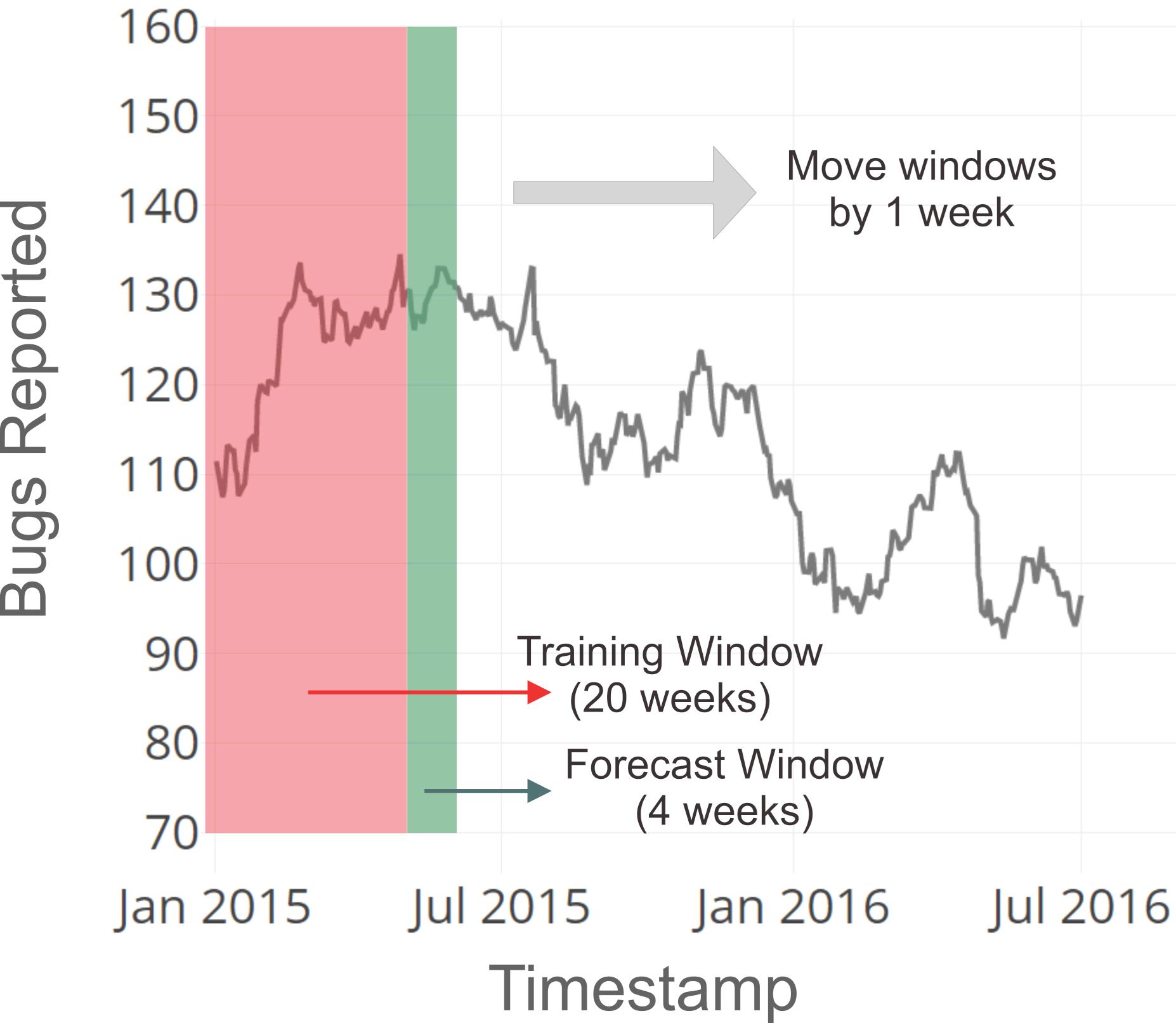}
	\caption{Rolling window approach for forecasting issues, bugs, and enhancements over time. The input window has a duration of 20 weeks ($\approx 4$ months) and the forecast window is 4 weeks ($\approx 1$ month). The step length of the moving window is 1 week.}
	\label{fig:rolling_window}
\end{figure}

\subsection{Proposed Forecasting Approach}
\label{sect:approach}
In evaluating the performance of time series modeling, we used a rolling 
window time series forecasting approach as shown in \fig{rolling_window}. This 
approach works as follows:
\be
\item First, we create two windows (labeled training window and testing window 
in \fig{rolling_window}). After extensive empirical evaluation of all the 
projects, we determined the best training window size to be around 20 weeks 
($\approx$ 4 months) and test window as set to 4 weeks ($\approx$ 1 month).
\item Next, we train an ARIMA model on the time series data from the training 
window and forecast for issues, bugs, and enhancements over the duration of the 
test window.
\item Then, we estimate the magnitude of average error (also known as MAE, 
described in \tion{stats}) of the forecast values.
\item Finally, we move the training and testing window by 1 time step (in our 
case this is 1 week) and repeat steps 1, 2, and 3 until we reach the end of the 
time series.
\ee

After the rolling widow approach described above terminates, we gather the MAE 
values and compute the Mean MAE values and the spread 
computed as the variance in of the MAE values. These values are 
plotted for all the projects as shown in Figures \ref{fig:rq1} and \ref{fig:rq3}.


\section{Experimental Results}
\label{sect:results}
\subsection*{\normalsize{RQ1: Are there temporal trends in our data?}}
				\begin{figure}[tp!]
	\centering
	\begin{subfigure}[t]{\linewidth}
	\begin{subfigure}[t]{0.33\linewidth}
		\centering
		\includegraphics[width=\linewidth]{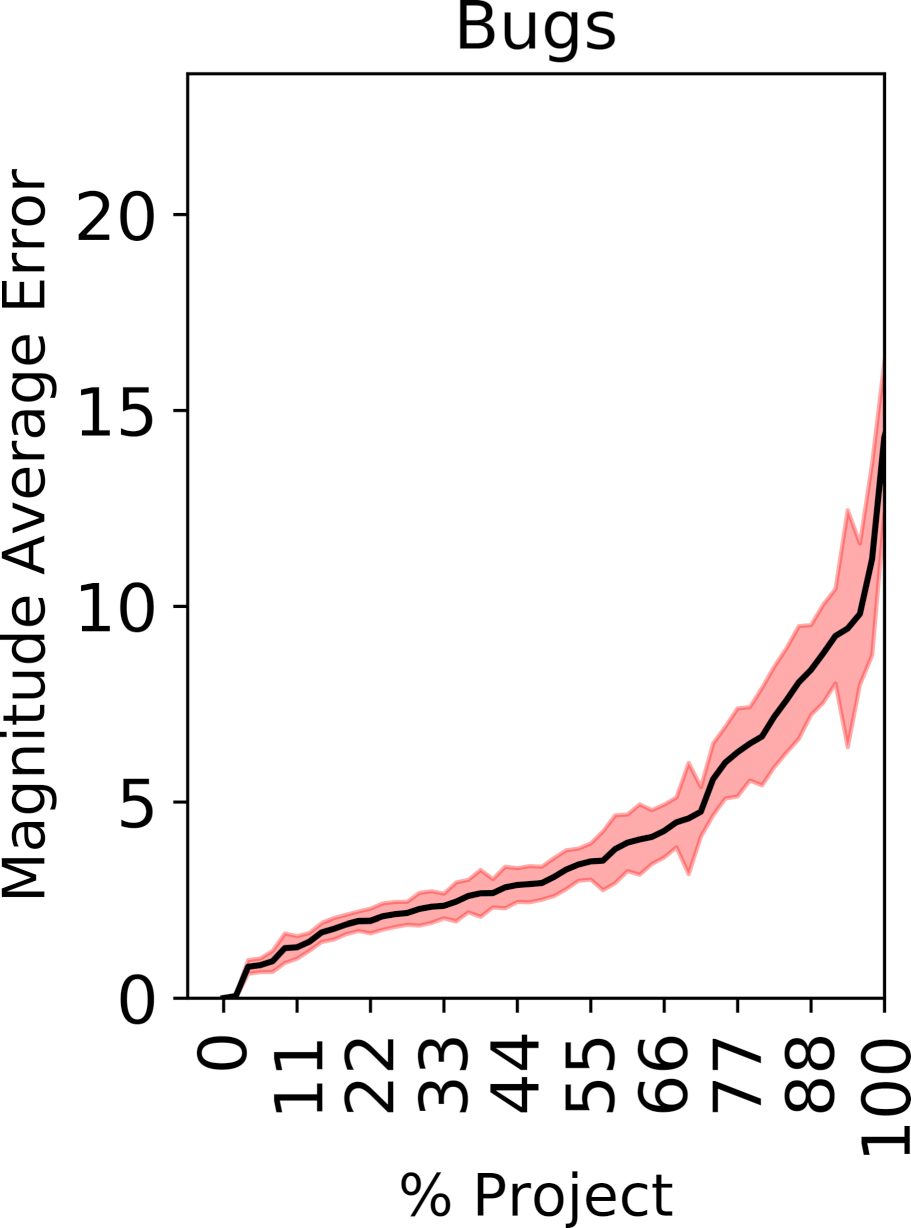}
	\end{subfigure}%
	~
	\centering
		\begin{subfigure}[t]{0.33\linewidth}
		\centering
		\includegraphics[width=\linewidth]{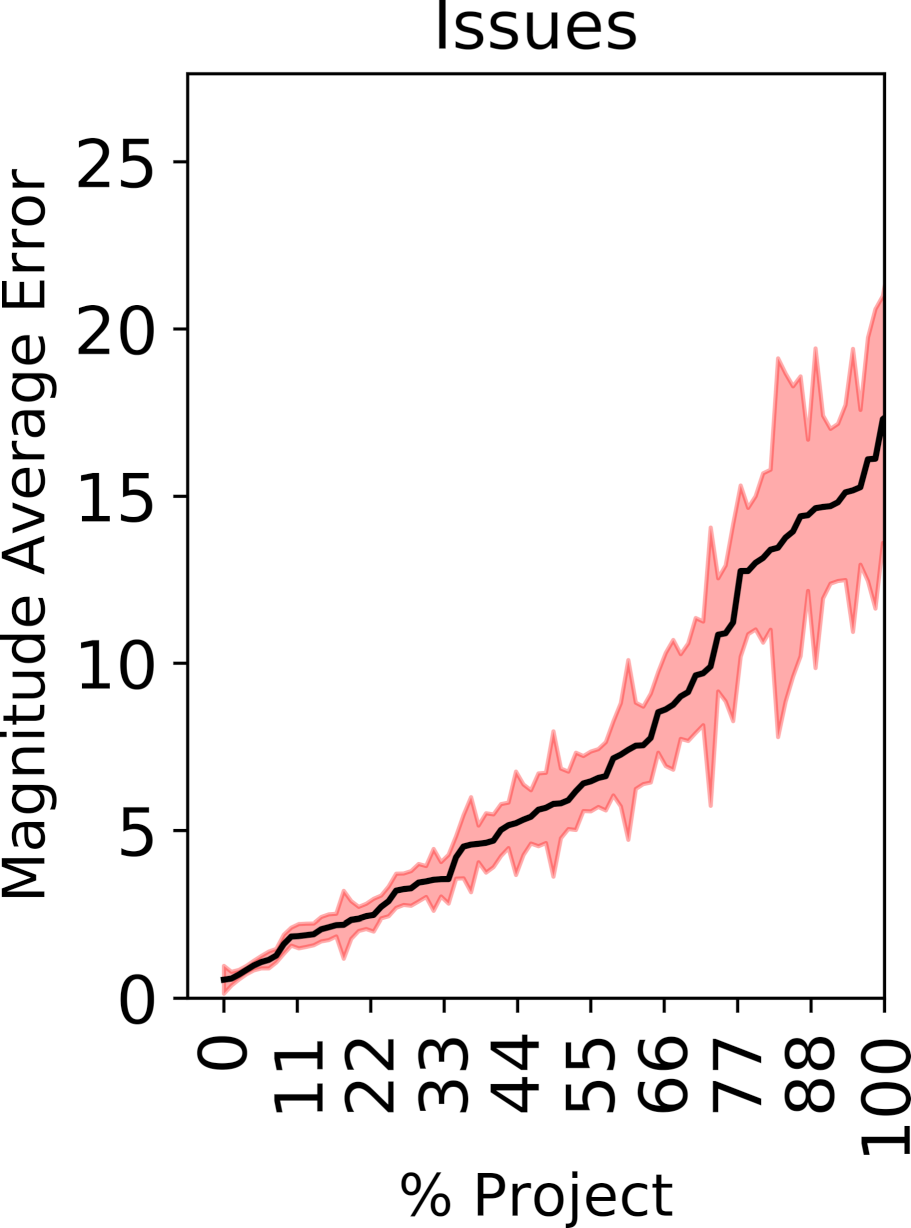}
	\end{subfigure}%
	~
	\centering
		\begin{subfigure}[t]{0.33\linewidth}
		\centering
		\includegraphics[width=\linewidth]{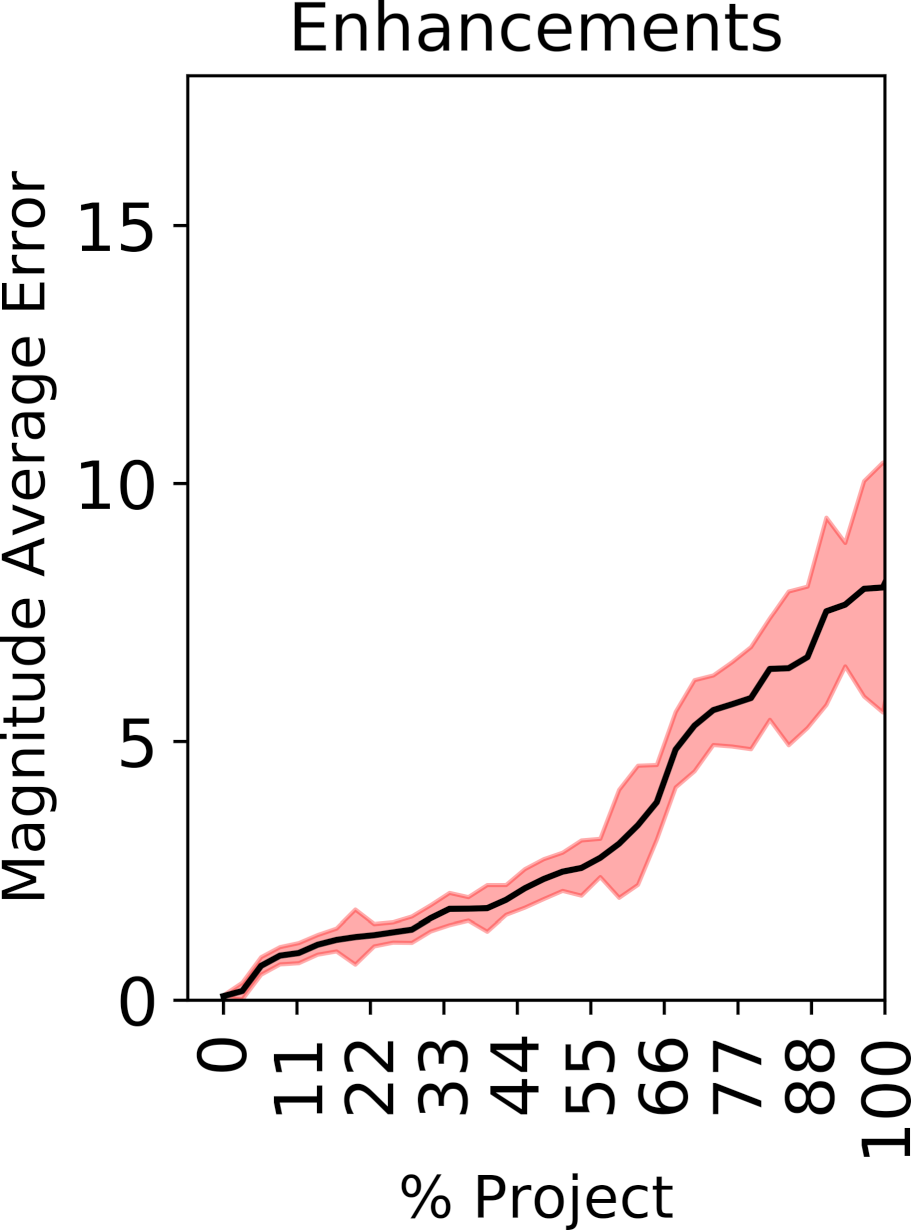}
	\end{subfigure}%
	\caption{171 Inhouse Datasets}
	\end{subfigure}
	\\
	\begin{subfigure}[t]{\linewidth}
	\begin{subfigure}[t]{0.33\linewidth}
	\centering
	\includegraphics[width=\linewidth]{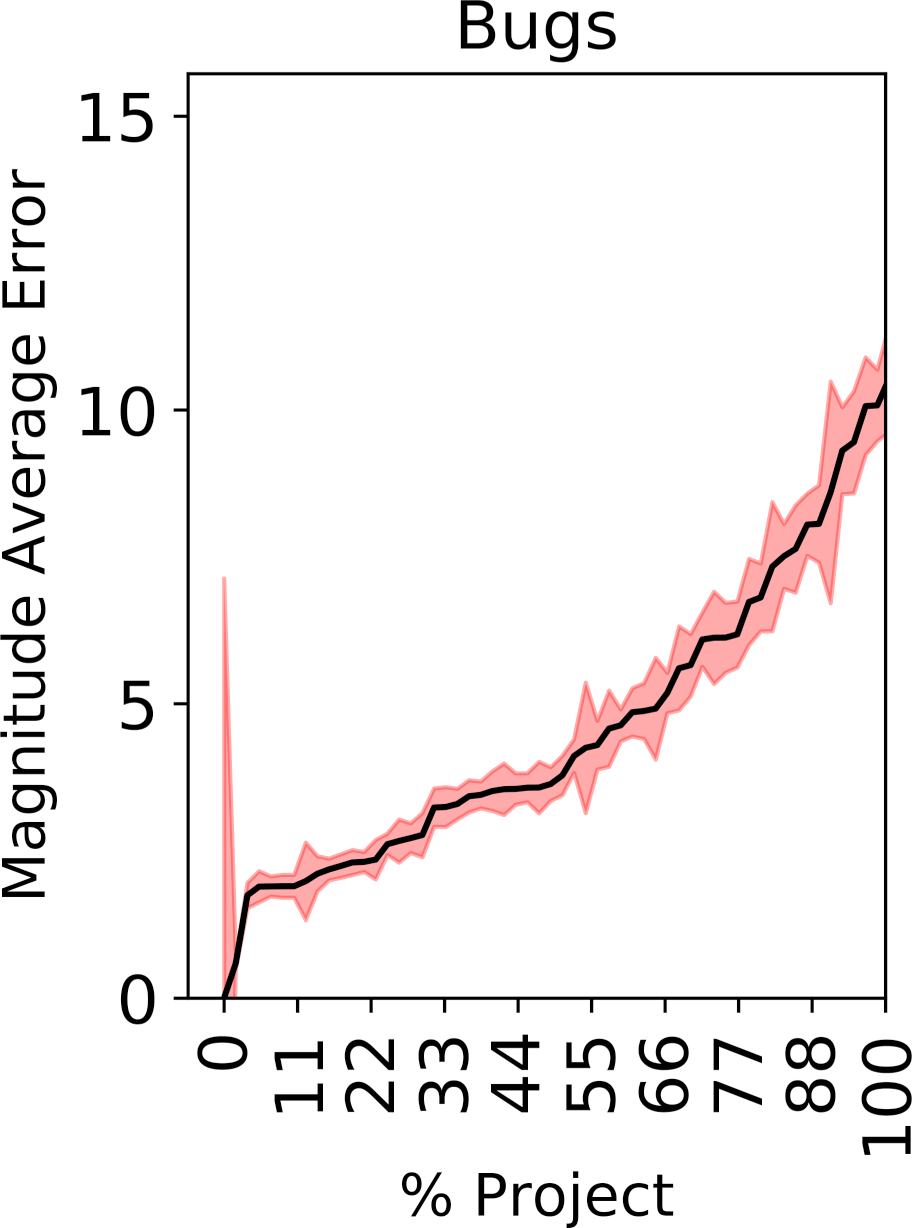}
	\end{subfigure}%
	~
	\centering
	\begin{subfigure}[t]{0.33\linewidth}
		\centering
		\includegraphics[width=\linewidth]{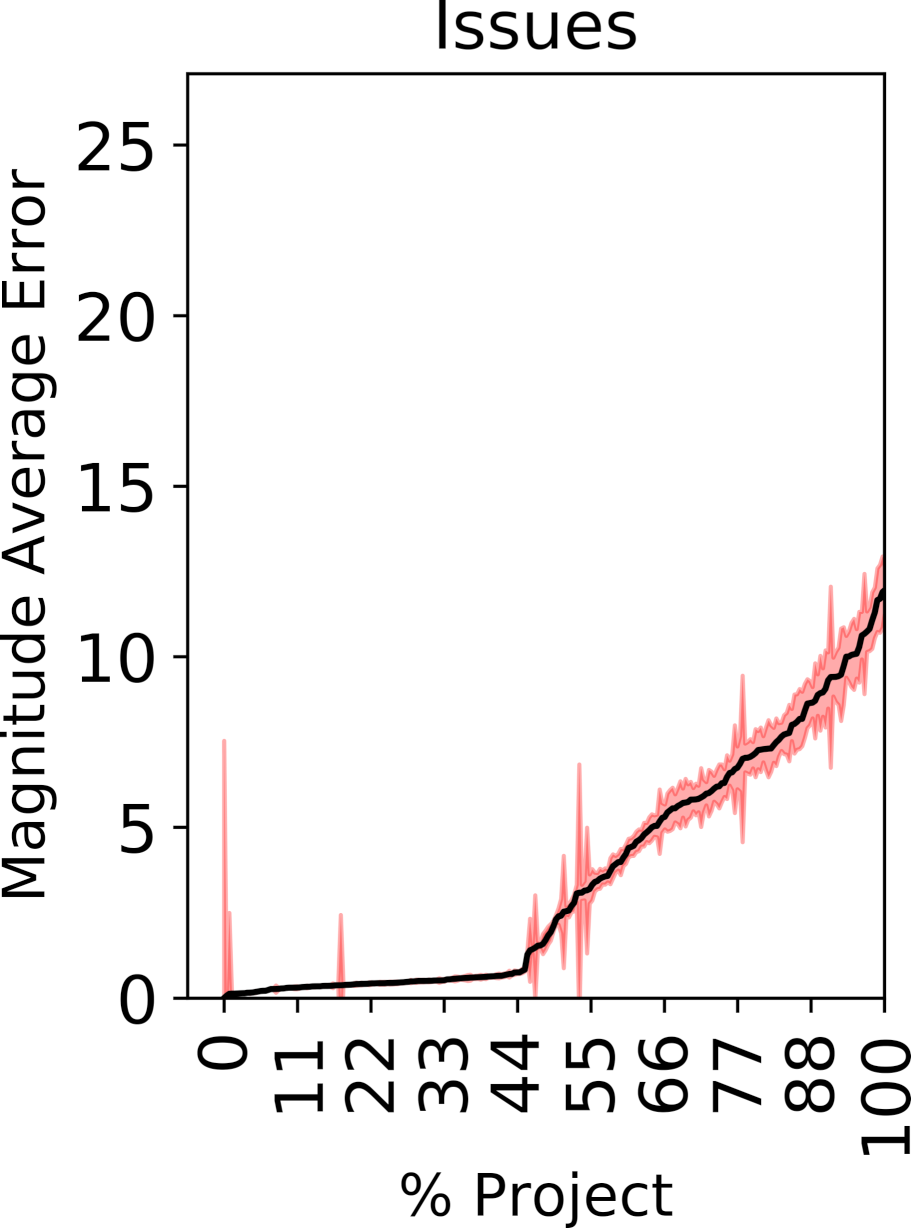}
	\end{subfigure}%
	~
	\centering
	\begin{subfigure}[t]{0.33\linewidth}
		\centering
		\includegraphics[width=\linewidth]{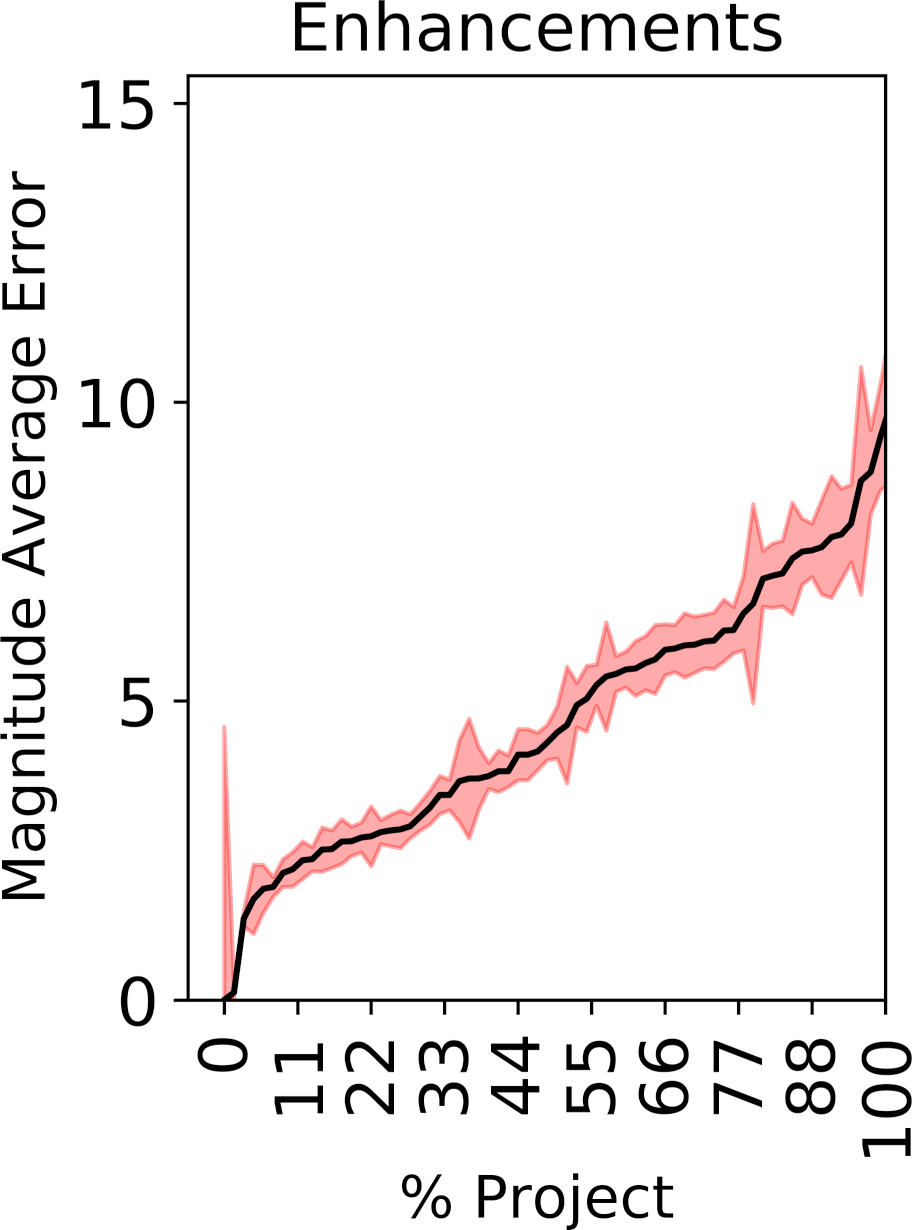}
	\end{subfigure}%
	\caption{671 Opensource datasets}
	\end{subfigure}
	\caption{This figure verifies the existence of temporal trends in issues, bugs, and enhancements. The low magnitude of average error (MAE) values show that time series forecasting with ARIMA can be performed on these attributes.}
	\label{fig:rq1}
\end{figure}

		The first research question seeks to establish the presence of temporal 
		trends in issues, bugs, and enhancements. If temporal trends do exist in 
		these attributes, a time series model such as ARIMA, which is equipped to 
		make accurate forecasts, should lead to low errors when we use past data to 
		forecast for the future. For this purpose, we ask if we may construct a 
		time series model using past data from the same attributes (called $\mathit{LOCAL}$). That is, we attempt 
		to: (a) forecast for future the number of bugs using past trends in bug 
		reports, (b) forecast for future the number of issue reports using past 
		trends in issue reports, and (c) forecast for future the number of 
		enhancements using past trends in enhancements.
			
		For experimentation, in each project we use a rolling window 
		method to train an ARIMA model on past 20 weeks and forecast for future 4 
		weeks. This is repeated for issues, bugs, and enhancements. Then, we 
		compute the magnitude of absolute error (MAE) between actual and forecast 
		values for each step of the rolling window.
		
		Our results are shown in~\fig{rq1}. For proprietary projects, in terms of 
		MAE, the errors for forecasting bugs in 66\% of the 
		projects are very small (they are close to zero in several cases). Further, we note that 
		the variance of these errors shaded in \colorbox{pink}{pink} in~\fig{rq1}
		are also quite low. For opensource projects, we notice similar trends. However, in case of open source projects, the MAE 
		scores are slightly higher when compared to proprietary projects. This means 
		that temporal trends do exist in open source projects, but these are less 
		temporal as compared to proprietary projects. In summary, we answer this 
		research question as follows:\\[-0.1cm]
		
		\noindent\begin{minipage}{\linewidth}
			\begin{center}
				\begin{tabular}{p{0.95\linewidth}}
					\arrayrulecolor{Gray}
					\hline
					\rowcolor{Gray}
					\textbf{\textit{Lesson 1}}\bigstrut\\
					\rowcolor{Gray} The mined attributes of both proprietary and opensource 
					projects exhibit temporal trends. Proprietary projects are slightly more temporal 
					compared to opensource projects.\bigstrut[b]\\\hline
				\end{tabular}
			\end{center}
		\end{minipage}\bigstrut[t]\\[-0.3cm]

\subsection*{\normalsize{RQ2: Are there correlations between mined attributes?}}
\begin{figure}[t!]
	\centering
	\begin{subfigure}[t]{0.45\linewidth}
		\centering
		\includegraphics[width=0.99\linewidth]{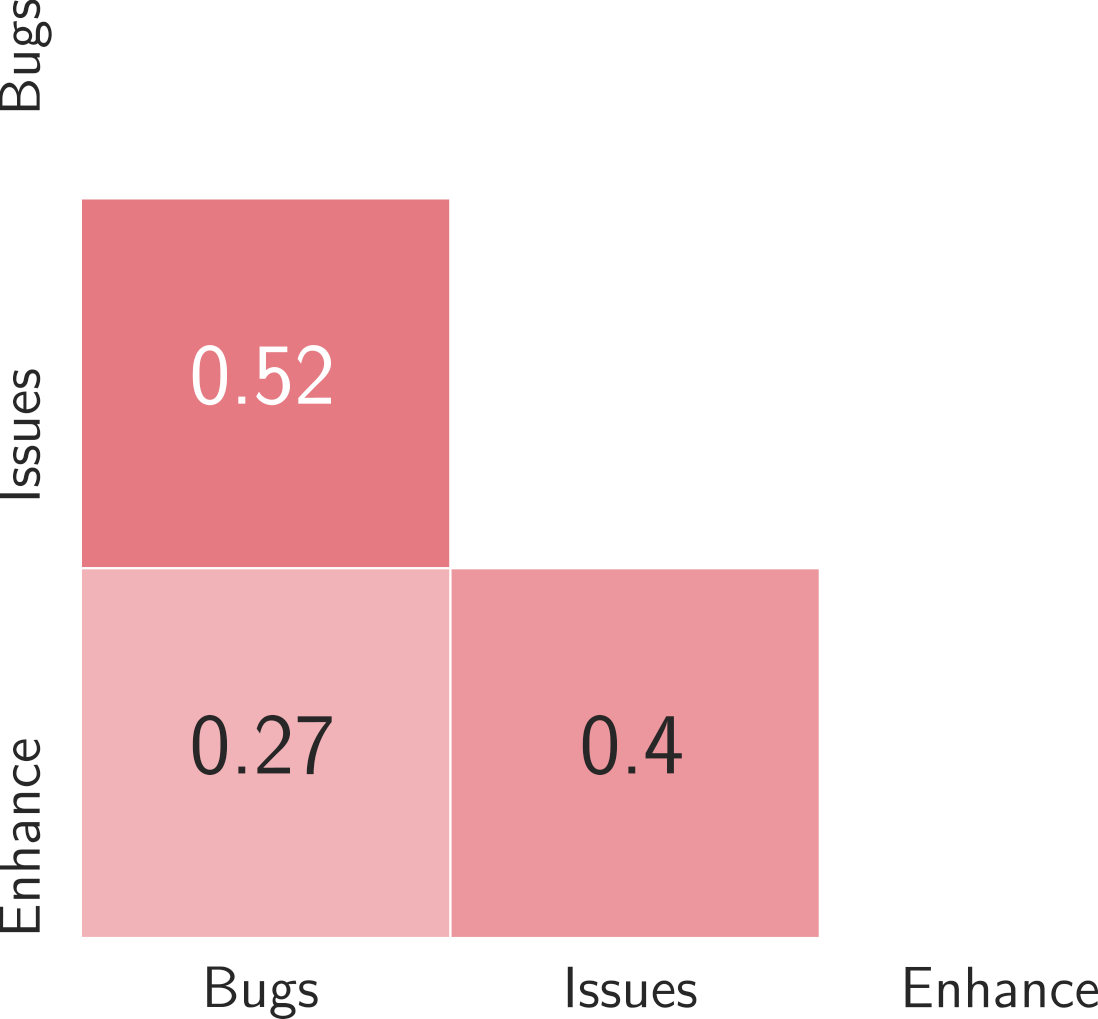}
		\caption{}
		\label{rq2:a}
	\end{subfigure}
	\begin{subfigure}[t]{0.45\linewidth}
	\centering
	\includegraphics[width=\linewidth]{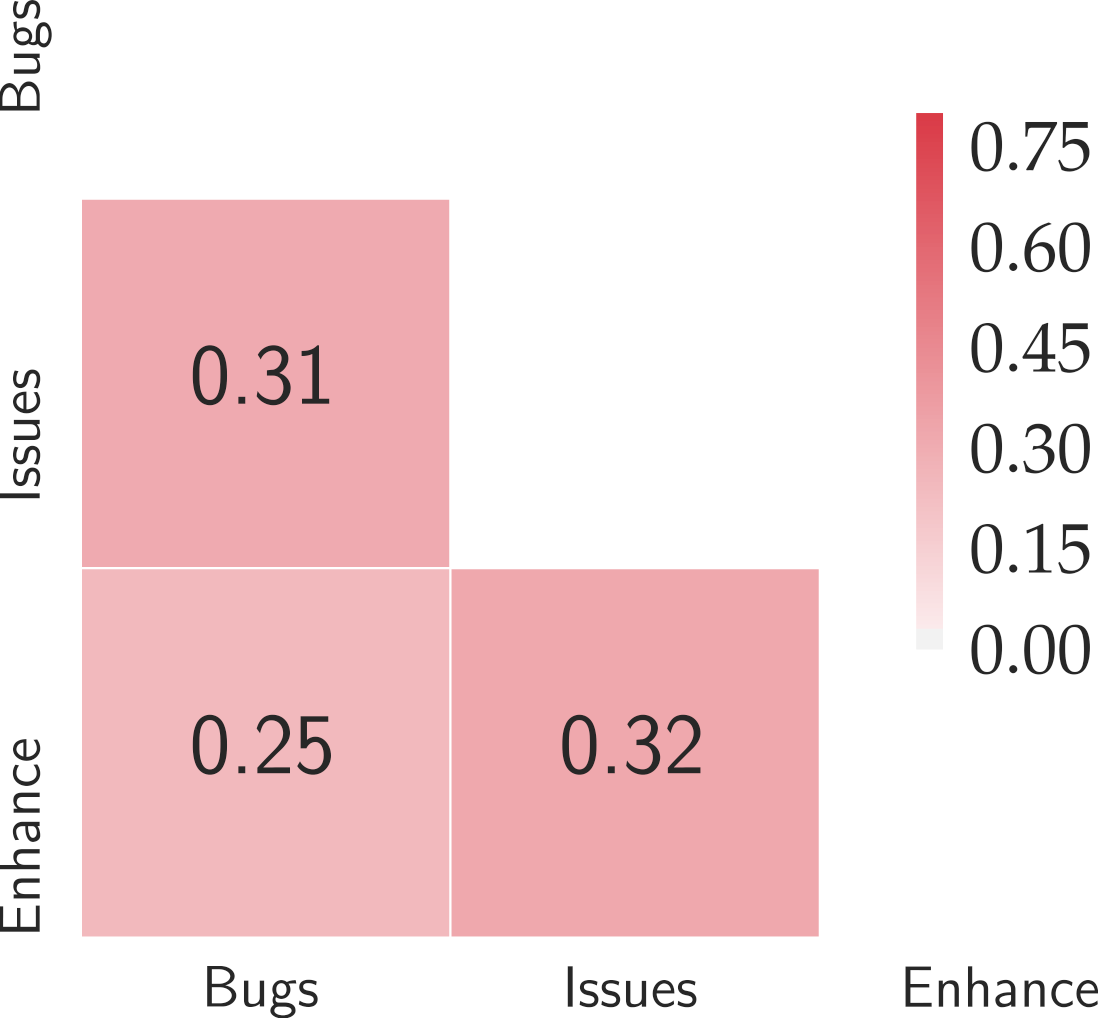}
		\caption{}	
		\label{rq2:b}	
	\end{subfigure}%
	\caption{Spearman's $\rho$. Figure~\ref{rq2:a} shows the correlations in 
	proprietary projects. Figure~\ref{rq2:b} shows the correlations in opensource 
	projects.}
	\label{fig:rq2}
\end{figure}

Having established that the attributes we mined (issues, bugs, and 
enhancements) exhibit temporal trends, in this research question we explore the 
correlations between mined attributes. This research question was partly 
motivated by the findings of Ayari et al.~\cite{ayari2007threats}, they report 
that,\textit{ ``\dots at least half of all 
the issues in a tracking system are 
related to bugs and the other half contains a mix of preventive, 
perfective and corrective requests.''} This work was published in 2007, and 
since then there has been a wide adoption of version control systems such as 
GitHub by several projects. These version control systems have integrated issue 
tracking mechanisms such as 
GitHub issues. Therefore, in this research question, we revisit this claim to 
check for the relationship between issues, bugs, and enhancements.

For each of our 832 proprietary and opensource projects, we compute the 
correlation between all pairs of attributes. For this, we used Spearman's 
$\rho$. That is, we compute correlations between:
\bi[leftmargin=*]
\item $issues \leftrightarrow bugs$  
\item $enhancements \leftrightarrow bugs$ 
\item $issues \leftrightarrow enhancements$ 
\ei

\begin{figure*}[htbp]
	\centering
	\textbf{Proprietary}\\[0.2cm]
	\begin{subfigure}[t]{0.49\linewidth}
		\centering
		\textbf{Bugs}\\[0.1cm]
		\includegraphics[width=0.335\linewidth]{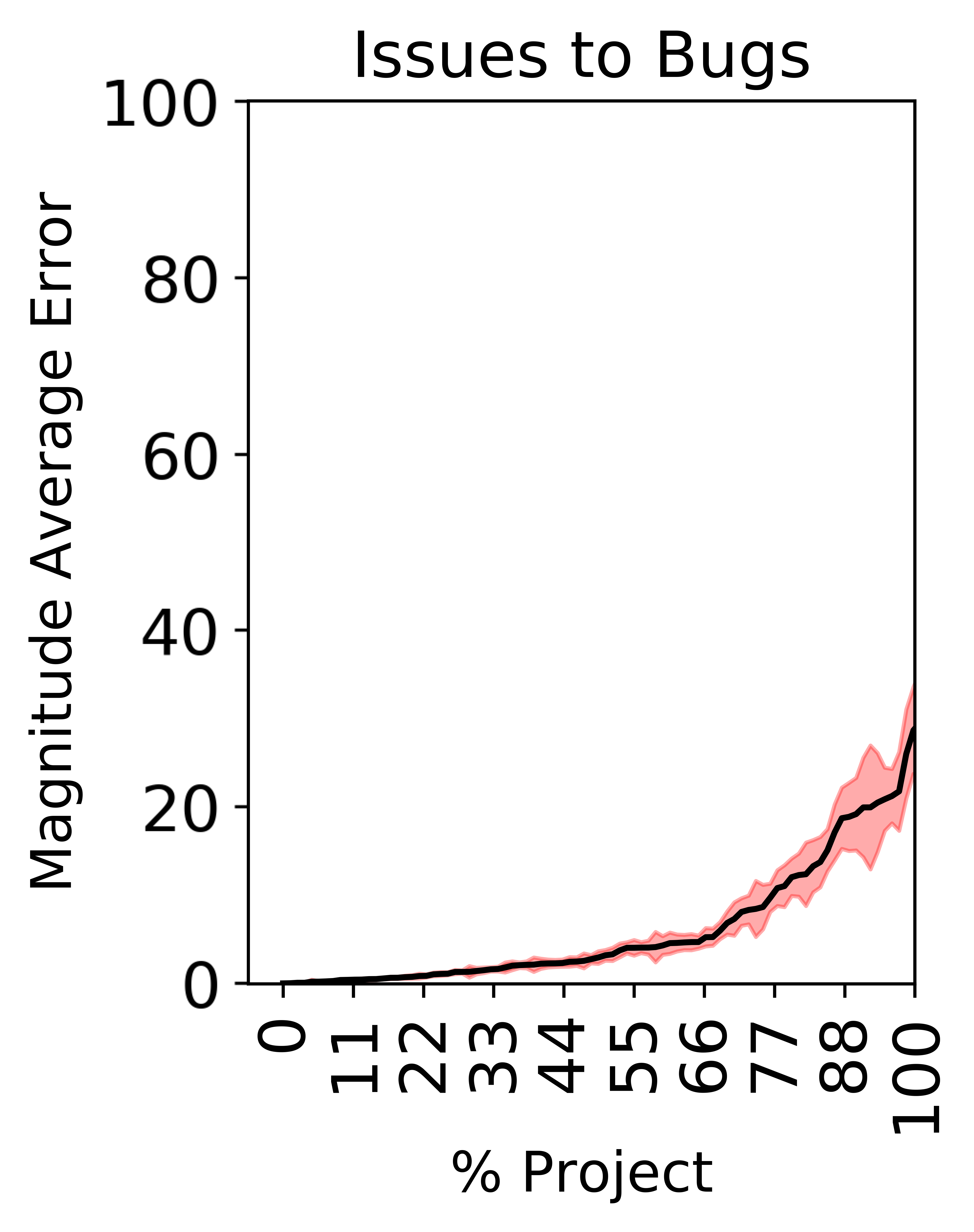}
		\includegraphics[width=0.33\linewidth]{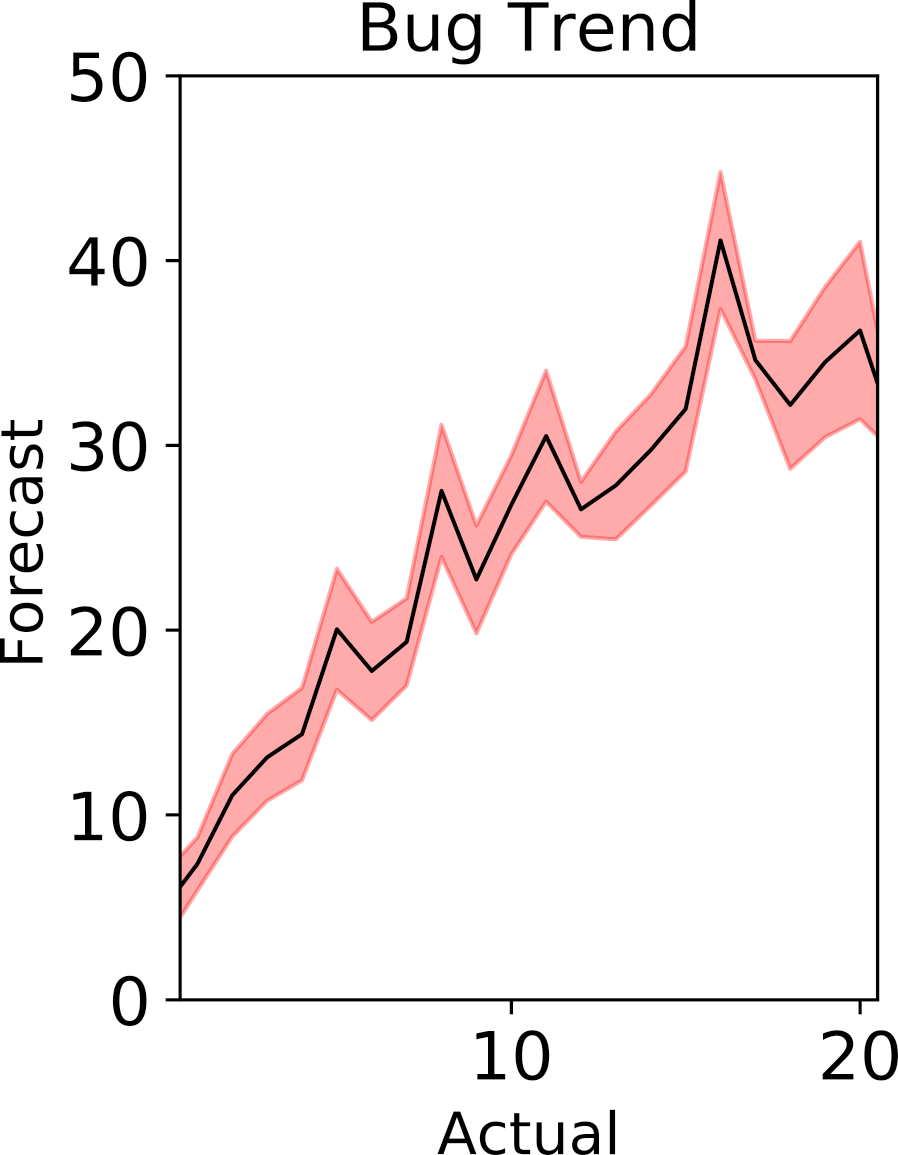}
\end{subfigure}%
	\centering
		\begin{subfigure}[t]{0.5\linewidth}
		\centering
		\textbf{Enhancements}\\[0.1cm]
		\includegraphics[width=0.334\linewidth]{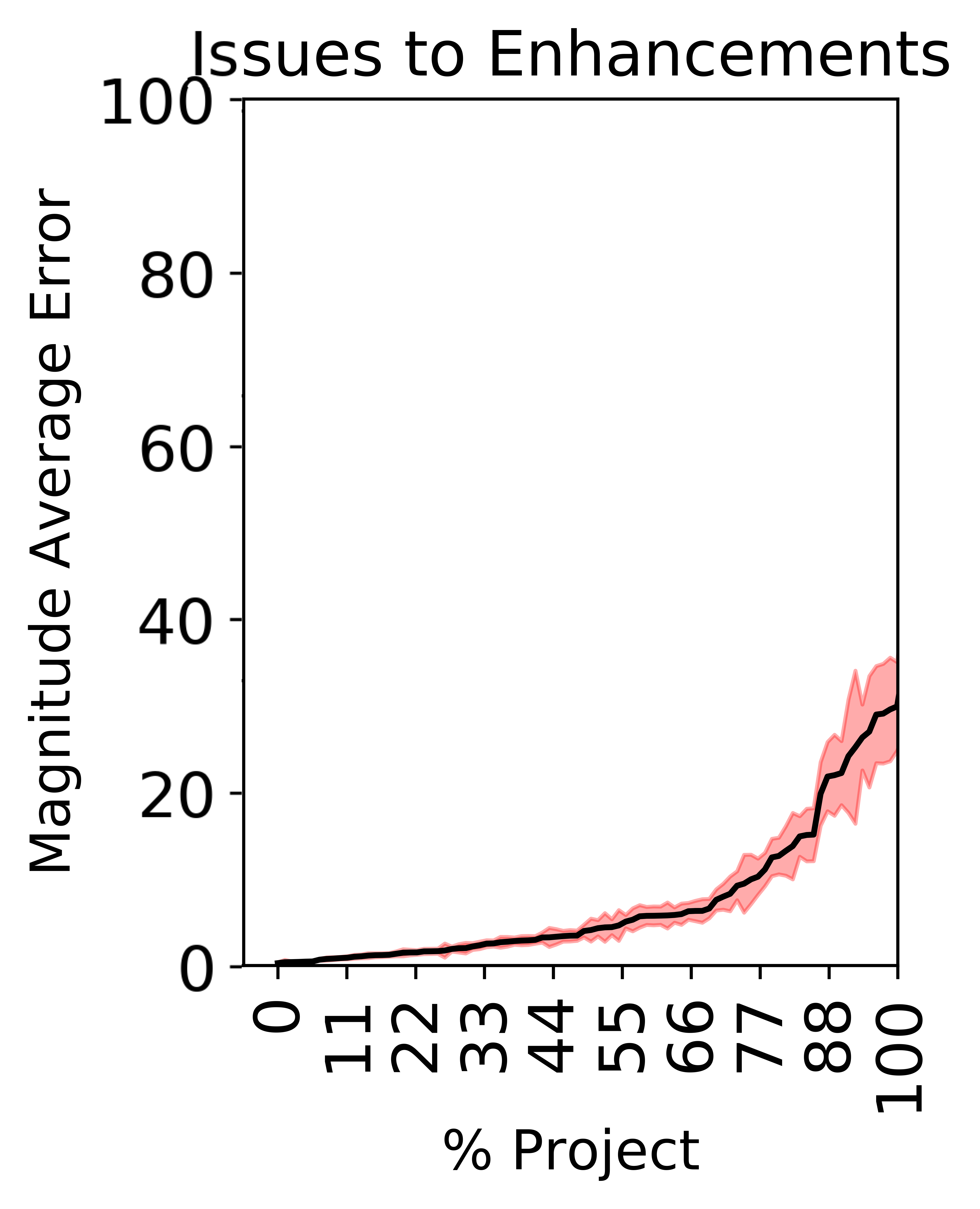}
		\includegraphics[width=0.32\linewidth]{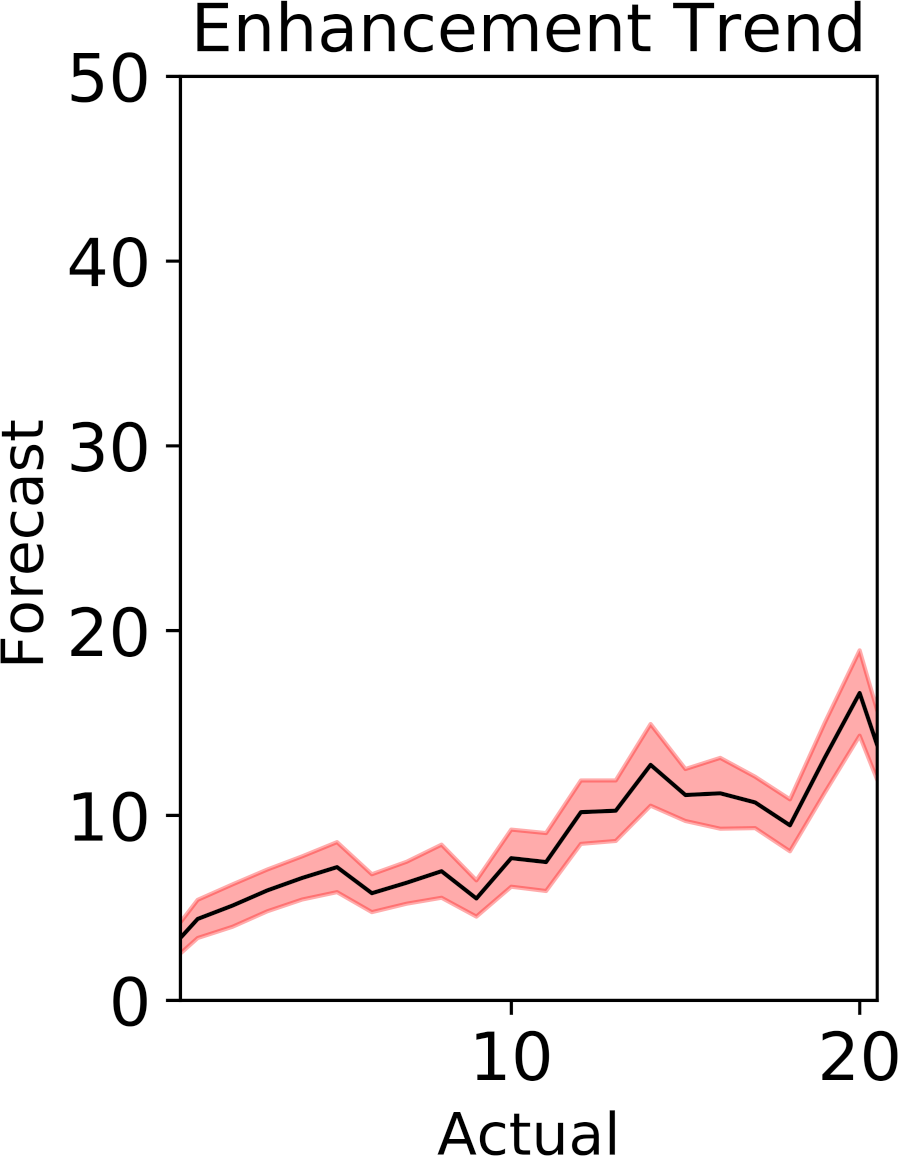}
	\end{subfigure}%
	\\[0.2cm]
	\textbf{Opensource}
	\\[0.1cm]
	\begin{subfigure}[t]{0.5\linewidth}
		\centering
		\includegraphics[width=0.335\linewidth]{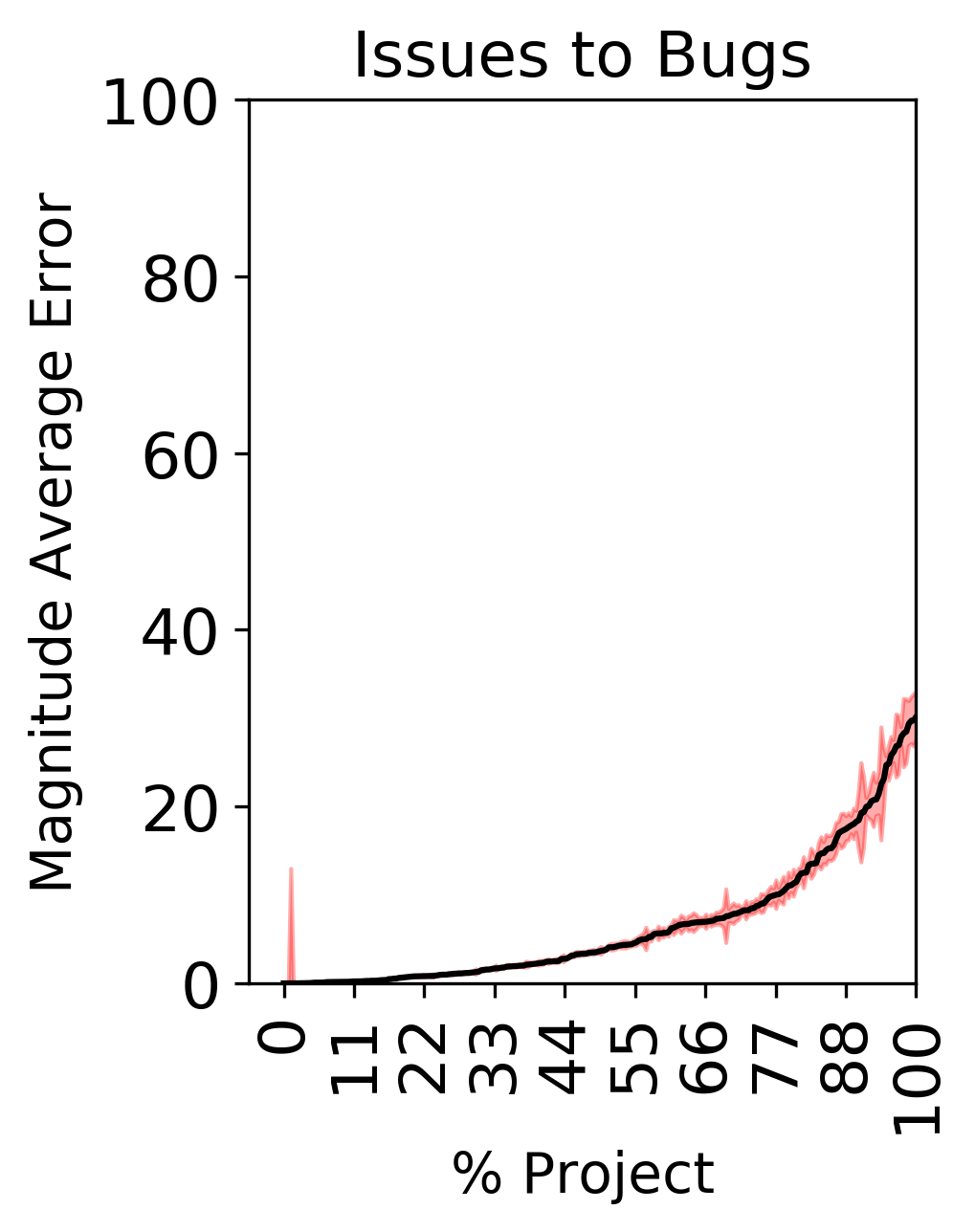}
		\includegraphics[width=0.33\linewidth]{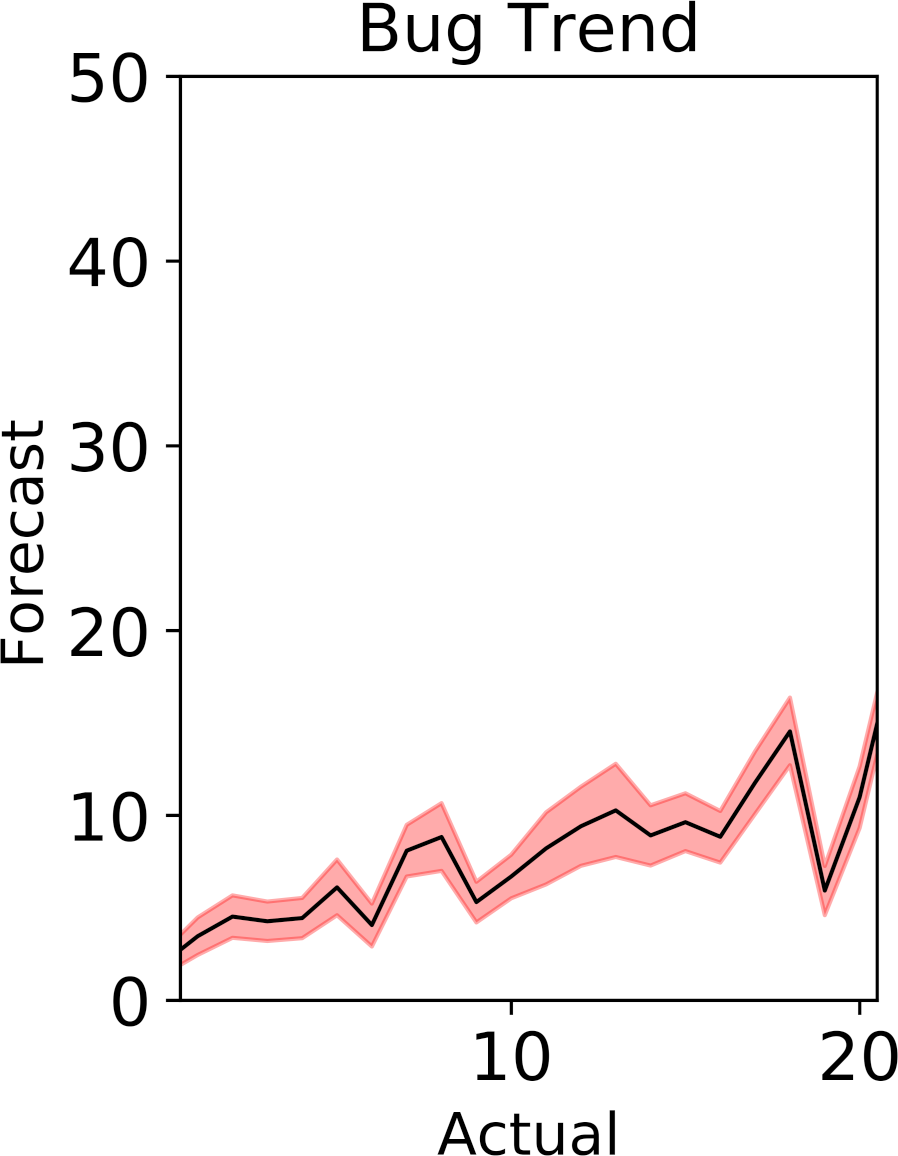}
	\end{subfigure}%
	\centering
		\begin{subfigure}[t]{0.5\linewidth}
		\centering
		\includegraphics[width=0.334\linewidth]{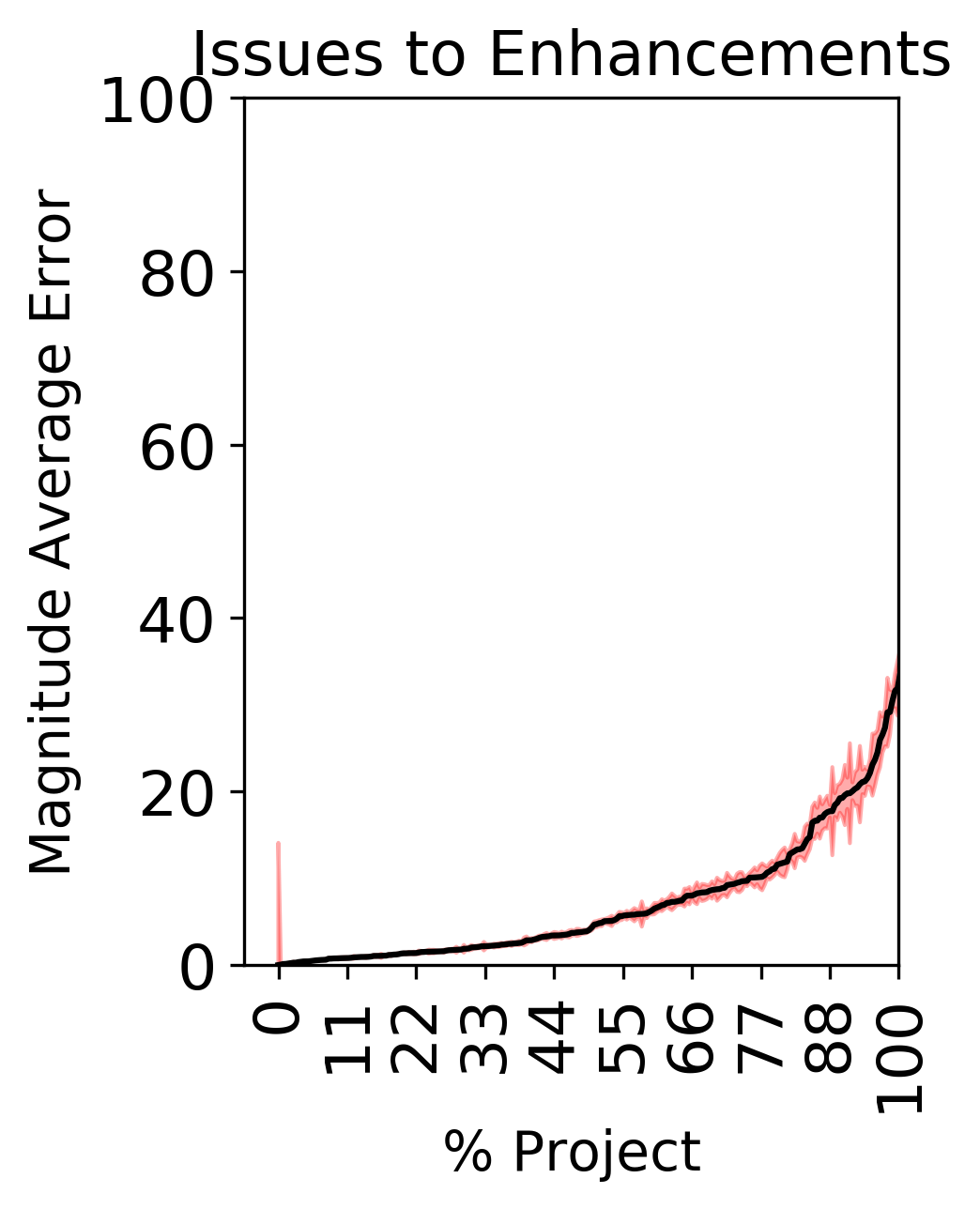}
		\includegraphics[width=0.32\linewidth]{images/RQ3/inhouse/enhancements/2.png}
	\end{subfigure}%
	\caption{This figure shows forecasts for future bugs and enhancements using ARIMA models constructed using past issue reports. For both proprietary and opensource projects, the magnitude of average error is very low (close to zero in several cases). Trend graphs show that an increase in actual bugs (or enhancements) leads to a corresponding increase in forecasts.}
	\label{fig:rq3}
\end{figure*}

\fig{rq2} shows a heatmap with the correlation values. In this figure, a value 
close to 1 would indicate a strong positive correlation, a value close to $-1$ 
would indicate a strong negative correlation, and a value close to 0 would 
indicate no correlation. A $\rho$ value between 0.3 and 0.7 is considered 
moderate to strong~\cite{seber2012linear}. Our findings corroborate the report 
of Ayari et al. We make the following observations:
\be
\item In proprietary projects, there exist two moderate to strong correlations: 
(a) Our strongest correlation exists between issues $\leftrightarrow$ bugs, and 
(b) This is followed by issues $\leftrightarrow$ enhancements.
\item In opensource projects, there still exists the same two correlations.  
But, these correlations are relatively weaker compared to same correlations in 
proprietary projects.
\ee

\noindent We summarize our findings as follows:\\[-.1cm]

\noindent\begin{minipage}{\linewidth}
	\begin{center}
		\begin{tabular}{p{0.95\linewidth}}
			\arrayrulecolor{Gray}
			\hline
			\rowcolor{Gray}
			\textbf{\textit{Lesson 2}}\bigstrut\\
			\rowcolor{Gray} There exist moderately strong correlations between 
$issues$ $\leftrightarrow$ $\{bugs , enhancements\}$. These correlations are 
stronger in proprietary projects compared to opensource 
projects.\bigstrut[b]\\\hline
\end{tabular}
\end{center}
\end{minipage}\bigstrut[t]\\[-0.2cm]

\subsection*{\normalsize{RQ3: Can issues forecast for future bugs and enhancements?}}

From RQ2, we learn that there exists correlations between \mbox{issues} 
$\leftrightarrow$ \mbox{\{bugs, enhancements\}}. In this 
research question, we seek to leverage those correlations to establish if it is 
possible to construct an ARIMA model on issues (labeled $\mathit{ISSUES}$) and 
use that model to forecast for bugs and enhancements.

For experimentation, we used a rolling window to create an ARIMA model on 
issues, then with that model we forecast for 
bugs and enhancements (this approach is described in detail in 
\tion{approach}). The use of rolling window results in forecasts for each time 
step progressed by the window. For these time steps, we measure the error 
between the forecast values bugs (and enhancements) and the actual values of 
bugs (and enhancements) using magnitude of average error (MAE, see \eq{mae}).
\fig{rq3} shows the magnitude of average error for bugs and 
enhancements in opensource and proprietary projects.

In order to see if the ARIMA model constructed using issues captures trends in 
bugs and enhancements, we plot the actual and 
forecast values of bugs (and enhancements) a function of bug (and 
enhancement) counts. We would expect an increase in the forecast of bug counts 
(or enhancement counts) as the actual values increase. These are shown 
in~\fig{rq3} as well.

Our findings are summarized as follows: 
\be
\item For most projects (both proprietary and opensource) the magnitude of 
average error very small. 
\item The magnitude of average errors in proprietary projects are slightly 
lower than in opensource projects.
\item The variance of these errors (which results from using the rolling 
window) are shaded in \colorbox{pink}{pink} are also noticeably low.
\ee
\noindent These results are very encouraging indeed and we answer this research 
question as follows:\\[0.1cm]
\noindent\begin{minipage}{\linewidth}
	\begin{center}
		\begin{tabular}{p{0.95\linewidth}}
			\arrayrulecolor{Gray}
			\hline
			\rowcolor{Gray}
			\textbf{\textit{Lesson 3}}\bigstrut\\
			\rowcolor{Gray} We find that ARIMA models built 
			on issues can be accurate for forecasting bugs and 
			enhancements for both proprietary and opensource projects. The errors 
			are very low (close to zero) 
			and the variance in errors are also significantly low.\\\hline
		\end{tabular}
	\end{center}
\end{minipage}\bigstrut[t]\\[-0.3cm]

\subsection*{\normalsize{RQ4: Are the forecasts using issues better than with past temporal data?}}
\begin{figure}[tb!]
	\centering
	\begin{subfigure}[t]{0.49\linewidth}
		\centering
		\includegraphics[width=0.5\linewidth]{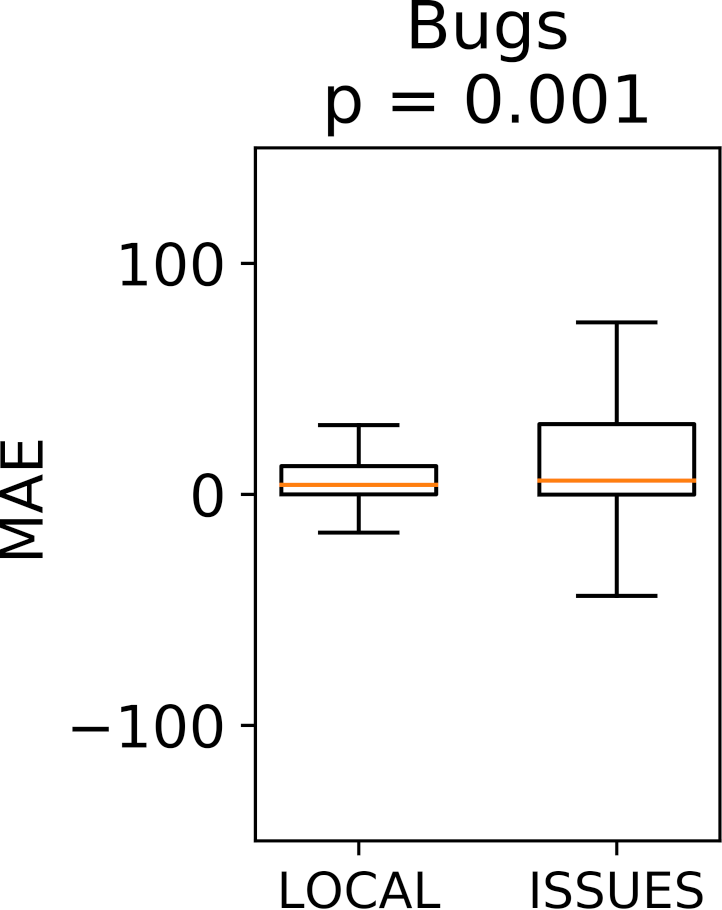}		
		\includegraphics[width=0.475\linewidth]{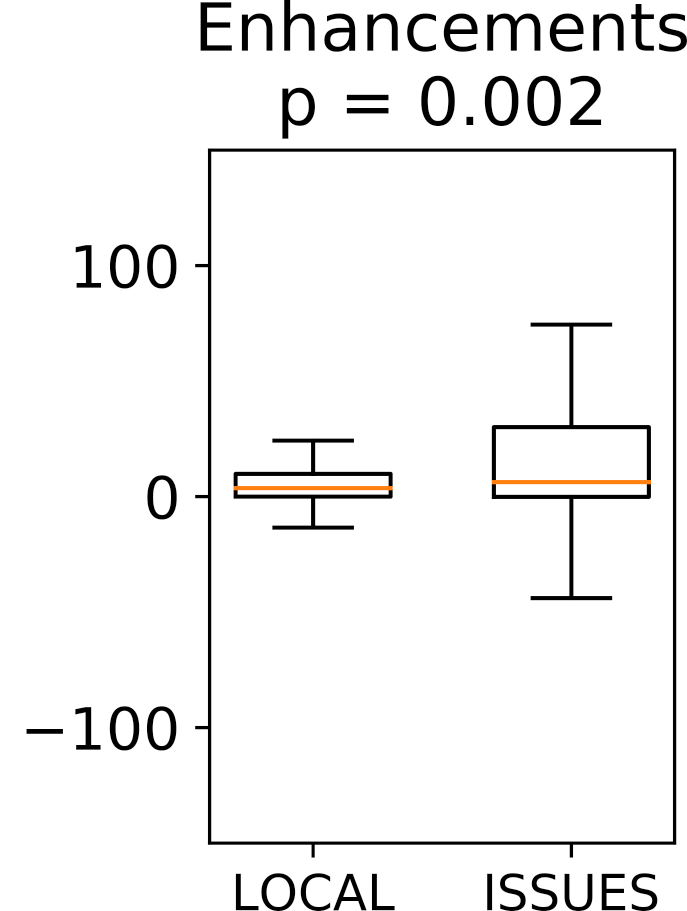}
		\caption{Proprietary}
	\end{subfigure}%
		\begin{subfigure}[t]{0.49\linewidth}
		\centering
		\includegraphics[width=0.5\linewidth]{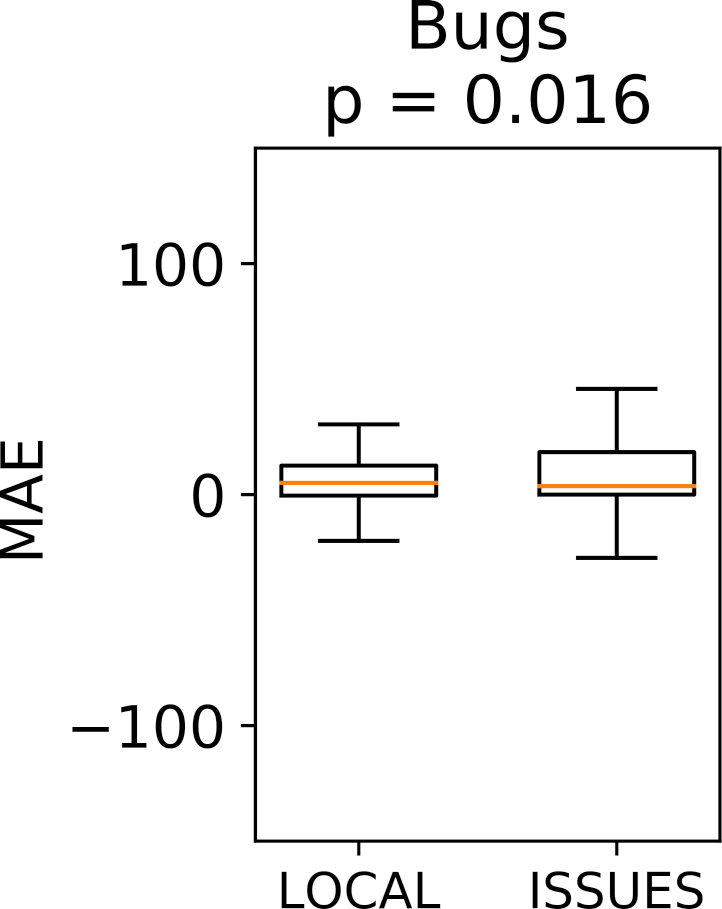}		
		\includegraphics[width=0.475\linewidth]{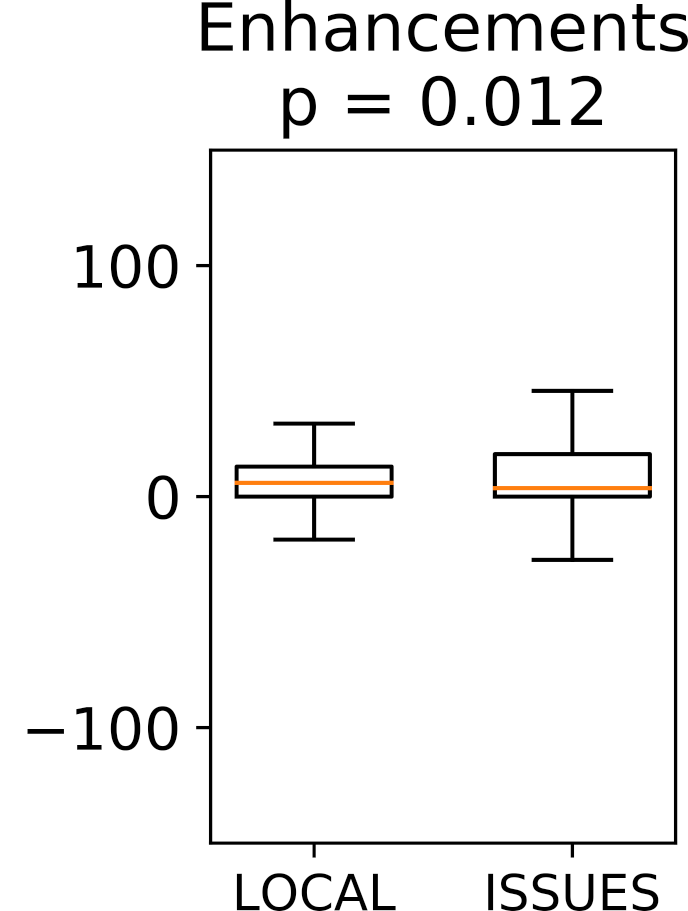}
		\caption{Opensource}
	\end{subfigure}%
	\caption{Compare distribution of errors with ARIMA forecasts using issue reports ($\mathit{ISSUE}$) and past temporal data ($\mathit{LOCAL}$). The charts indicate that the distribution of errors are very similar to each other with p-values $<$ 0.05.}
	\label{fig:rq4}
\end{figure}

In RQ1, we established that ARIMA model built on past bug and enhancement data (called $\mathit{LOCAL}$) can forecast for future bugs and enhancements with very low errors (see~\fig{rq1}). Further, in RQ3, we showed that an ARIMA model can built on past issue data (called $\mathit{ISSUE}$) can also forecast for future bugs and enhancements with low magnitudes of average errors (see~\fig{rq3}). This research question is a natural extension of those two results. Here we compare errors in forecasts obtained from $\mathit{ISSUE}$ and $\mathit{LOCAL}$ where: 
\be
\item[] a) $\mathit{ISSUE}$ : ARIMA model built using past issue report trend.
\item[] b) $\mathit{LOCAL}$ : ARIMA model built using past bug report trends 
and past enhancement request trends.
\ee

To perform this comparison, we use both of these models to forecast for future bugs and enhancements. 
The construction of $\mathit{LOCAL}$ model is described RQ1 and RQ3 describes 
the construction of $\mathit{ISSUE}$. We compute the error in forecasts with  
$\mathit{ISSUE}$ and $\mathit{LOCAL}$ and use a parametric statistical hypothesis test 
(Welch's t-test) to compare the errors\footnote{We use a parametric test 
because it is known that the errors of ARIMA models have a normal 
distribution~\cite{box2015time}}. In order to conduct the hypothesis test we use the following hypothesis:
\begin{quote}
$\mathcal{H}$: The distributions of errors in using $\mathit{ISSUE}$ is significantly larger
than errors in using $\mathit{LOCAL}$.
\end{quote}

If the $p~value$ of the above hypothesis is less than 0.05, then we may reject 
that hypothesis and assert that, \textit{``the distributions of errors in using $\mathit{ISSUE}$ is statistically similar to errors in using $\mathit{LOCAL}$.''}

The distribution of errors for forecasting bugs and enhancements in proprietary and opensource projects are shown \fig{rq4}. It is immediately noticeable the expected value of the errors are close to zero in both $\mathit{ISSUE}$ and $\mathit{LOCAL}$ for all cases. Additionally, the $p~values$ are always less than 0.05 in all the cases. Therefore, the answer to this research question is:\\[0.1cm]
\noindent\begin{minipage}{\linewidth}
	\begin{center}
		\begin{tabular}{p{0.95\linewidth}}
			\arrayrulecolor{Gray}
			\hline
			\rowcolor{Gray}
			\textbf{\textit{Lesson 4}}\bigstrut\\
			\rowcolor{Gray} In the 832 projects studied here, forecasts made using past 
			temporal data 
		statistically comparable to	forecasts made using only the issue reports.\\\hline
		\end{tabular}
	\end{center}
\end{minipage}\bigstrut[t]\\

Note that this is a result of much practical
importance since the effort involved in building the issue models of {\bf RQ3} is much lower than the bug (and enhancement) models seen in {\bf RQ1}.  The comparable errors indicate that it is not necessary to identify bugs or enhancements separately from issues to forecast for their future values. Rather, we may simply mine for issues and use that to forecast for future bugs and enhancements. Doing this significantly reduces effort required to mine for each one individually. 



\section{Threats to Validity}
\label{sect:threats}
This work, like any other empirical study, is subject to the following threats 
to validity.
    
\be[leftmargin=-1pt]
\item[] \textit{Model Bias:} For building the time series models, in this
study, we elected to use ARIMA. We chose this method
because past studies shows that, for time series modeling, the 
results were superior to other method. Other time series modeling
methods like long short term memory (LSTM) models
have shown promise in other areas. That comparison is beyond the scope
of this paper.

\item[] \textit{Evaluation Bias:} 
This paper uses one measure of error, MAE (see \eq{mae}). 
Other quality measures often used in software engineering to quantify
the effectiveness of forecast. A comprehensive analysis using these measures
is left for future work. 
    
\item[] \textit{Sample bias:}
Our data was gathered by mining GitHub. Several data sets may often be noisy and 
uninteresting. To address this, we take the following steps: (1) Apply sanity checks to filter out irrelevant projects (\tion{datasets});
(2) Include a wide variety of projects; (3) Report mean and standard errors in all our measurements; and  
(4) Perform statistically sound comparisons when appropriate (e.g. in RQ4 we use Welch's t-test).

Further, we have discussed our findings with business users who are software engineers and managers at IBM. They agreed with the
general direction of findings: they stated that issues can be used
as indicators of future bugs and enhancements. They also
agreed that there are differences between open source and
proprietary software development, and we should not assume
these tools and techniques will help the practitioners of interest,
in the same manner.
\ee
\section{Discussion}
\label{sect:discuss}
\begin{figure}[b!]
\centering
\includegraphics[width=0.7\linewidth]{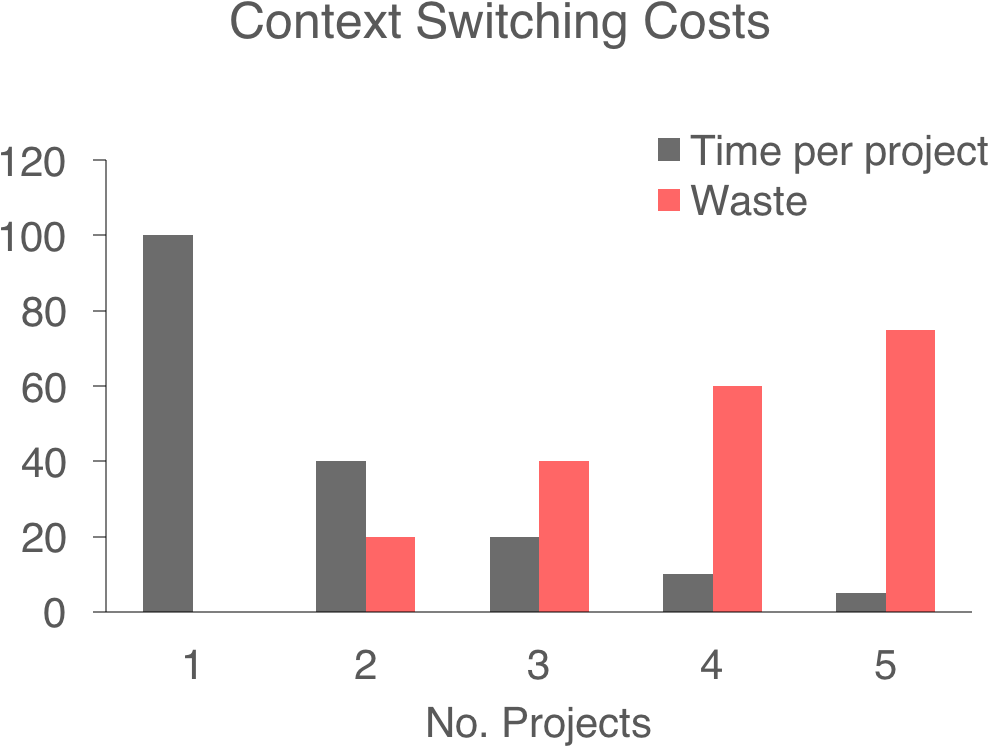}
\caption{Cost of switching contexts. Image courtesy 
of~\cite{weinberg1992quality}.}
\label{fig:context}
\end{figure}

A key motivation of this work is to reduce the amount of unexpected work assigned to any developer in any month.
With Agile models, developers are frequently reassigned to different projects. It is well established that a person who works on more than one project at a time incurs a cost in terms of time required to change contexts at each shift from one project to the other. The more complex the task, the more time it takes to make the shift~\cite{tugend_2008}. Gerald Weinberg~\cite{weinberg1992quality} showed that for software 
projects, the cost of switching between projects escalated if each task has a 
even a 10\% penalty as shown in~\fig{context}. In real world, the 
costs are usually much higher.

For large organizations that maintain their software suites based on end-user issue reports, 
it is vitally important for managers not to overload staff with tasks (since 
such overloading can lead to greatly reduced productivity). Hence, rather than
react to sudden changes in staffing level, a better approach is to plan weeks 
ahead in order to even out the workflow amongst all developers and alleviate 
effort involved in switching contexts. Of course, for that to work, we need a 
planning agent that can forecast the future. The results of this paper enables
this kind of reasoning to address the above managerial issue. Specifically, we
show that --- 
\begin{quote}
\textit{Managers need to only track issue trends in projects. With these trends 
they may be able to forecast the number of future bugs reports and enhancement requests.}
\end{quote}

We note that, this result would very useful for large service organizations maintaining suites 
of software attempting to effectively manage personnel across projects. 


\section{Conclusion}
\label{sect:conclusion}
In summary, we mined 832 projects (661 opensource and 171 proprietary projects) for issue reports over time. Following this, we spent over two months curating these issues into bugs and enhancements. The effort involved in the curation process lead us to investigate if these attributes are correlated with each other. We discovered that the attributes were indeed correlated and that it is possible to build time series models on one attribute (issues) and use that to forecast for bugs and enhancements. Our method can be used to circumvent much of the complex machinery required to commission and maintain a convention bug predictors. In addition to the simplicity, we show that our method is quite accurate (with near zero errors) in more than 66\% of the projects explored here. 

Our claim is not that issues provide the best forecast for bugs (or enhancements), rather it is that issue trends may be leveraged to supplement bugs forecast with a sufficiently high degree of accuracy. And these forecasts may be very useful for anticipating the required managerial actions. 

While our result are biased by our sample of projects, we have made attempts to include a wide array of projects both proprietary and opensource. To the
best of our knowledge, there exists no large study that reports the
opposite of our conclusions. At the very least, with our work we seek to highlight the merits of mining software repositories to perform time series analysis -– an issue that, we hope, is addressed further by other researchers on other projects.

\balance
\bibliographystyle{ACM-Reference-Format}
\bibliography{manuscript}

\end{document}